\documentclass[superscriptaddress,twocolumn,longbibliography,amsmath,amssymb,floatfix,aps]{revtex4-1}
\usepackage[]{graphicx,color}
\usepackage{amsmath}
\usepackage{algorithm}
\usepackage{algcompatible}
\usepackage{color}
\usepackage[utf8]{inputenc}
\usepackage{CJKutf8}
\usepackage{xspace}

\definecolor{darkgreen}{rgb}{0,0.5,0}

\begin{document}
\title{
Bayesian optimization of chemical composition:
a comprehensive framework
and its application to $R$Fe$_{12}$-type magnet compounds
}
\author{Taro \surname{Fukazawa}}
\email[E-mail: ]{taro.fukazawa@aist.go.jp}
\affiliation{CD-FMat, National Institute of Advanced Industrial Science
and Technology, Tsukuba, Ibaraki 305-8568, Japan}
\affiliation{ESICMM, National Institute for Materials Science,
Tsukuba, Ibaraki 305-0047, Japan}
\author{Yosuke \surname{Harashima}}
\affiliation{CD-FMat, National Institute of Advanced Industrial Science
and Technology, Tsukuba, Ibaraki 305-8568, Japan}
\affiliation{ESICMM, National Institute for Materials Science,
Tsukuba, Ibaraki 305-0047, Japan}
\author{Zhufeng \surname{Hou}}
\affiliation{CMI$^2$, National Institute for Materials Science,
Tsukuba, Ibaraki 305-0047, Japan}
\author{Takashi \surname{Miyake}}
\affiliation{CD-FMat, National Institute of Advanced Industrial Science
and Technology, Tsukuba, Ibaraki 305-8568, Japan}
\affiliation{ESICMM, National Institute for Materials Science,
Tsukuba, Ibaraki 305-0047, Japan}
\affiliation{CMI$^2$, National Institute for Materials Science,
Tsukuba, Ibaraki 305-0047, Japan}

\date{\today}
\begin{abstract}
We propose a
framework for
optimization of
the chemical composition of multinary compounds
with the aid of 
machine learning.
The scheme is based on first-principles calculation using the Korringa-Kohn-Rostoker method 
and the coherent potential approximation (KKR-CPA). 
We introduce a method for integrating datasets
to reduce systematic errors in a dataset, where 
the data are corrected using a smaller and more accurate dataset.
We apply this method to values of
the formation energy
calculated by KKR-CPA
for nonstoichiometric systems 
to improve them using a small dataset for stoichiometric systems
obtained by the projector-augmented-wave (PAW) method. 
We apply our framework to optimization of
$R$Fe$_{12}$-type magnet compounds
(R$_{1-\alpha}$Z$_{\alpha}$)(Fe$_{1-\beta}$Co$_{\beta}$)$_{12-\gamma}$Ti$_{\gamma}$, and 
benchmark the efficiency in
determination of the optimal choice of elements (R and Z) and ratio ($\alpha$, $\beta$ and $\gamma$) 
with respect to magnetization, Curie temperature and formation energy. 
We find that the optimization efficiency depends on descriptors significantly.
The variable $\beta$, $\gamma$ and the number of electrons from
the R and Z elements per cell are important in improving the efficiency.
When the descriptor is appropriately chosen, the Bayesian optimization becomes 
much more efficient than random sampling.
\end{abstract}
\preprint{Ver. 0.8.6}
\maketitle

\section{Introduction}
 Machine learning is attracting much attention these days,
 and its application to data obtained by first-principles calculation
 is a promising way to accelerate the exploration of
 novel materials.
 The basic idea
 is as follows:
 (i) introduce a numerical representation $x$ for the materials,
 which is called descriptor,
 (ii) calculate a property $y$ for materials from the search space
 of the descriptor by first-principles calculation, 
 and (iii) infer a relation $y=f(x)$ between $x$ and $y$
 from thus obtained data
 by modeling $f$.
 Many efforts have been made to develop models and descriptors that work
 in materials discovery.\cite{Rupp12,Ghiringhelli15,Seko15,Pham16,Pham17,Dam18,Pham18,Ouyang18,Seko18}
 These models can be used to identify promising candidates
 by predicting the property $f(x')$ for unknown materials $x'$.
 It is also possible to perform the modeling and the sampling alternately
 to obtain the optimal $x$ as quickly
 as possible, which is called optimization.

\begin{figure}
\centering
\includegraphics[bb=0 0 504 360, width=9cm]{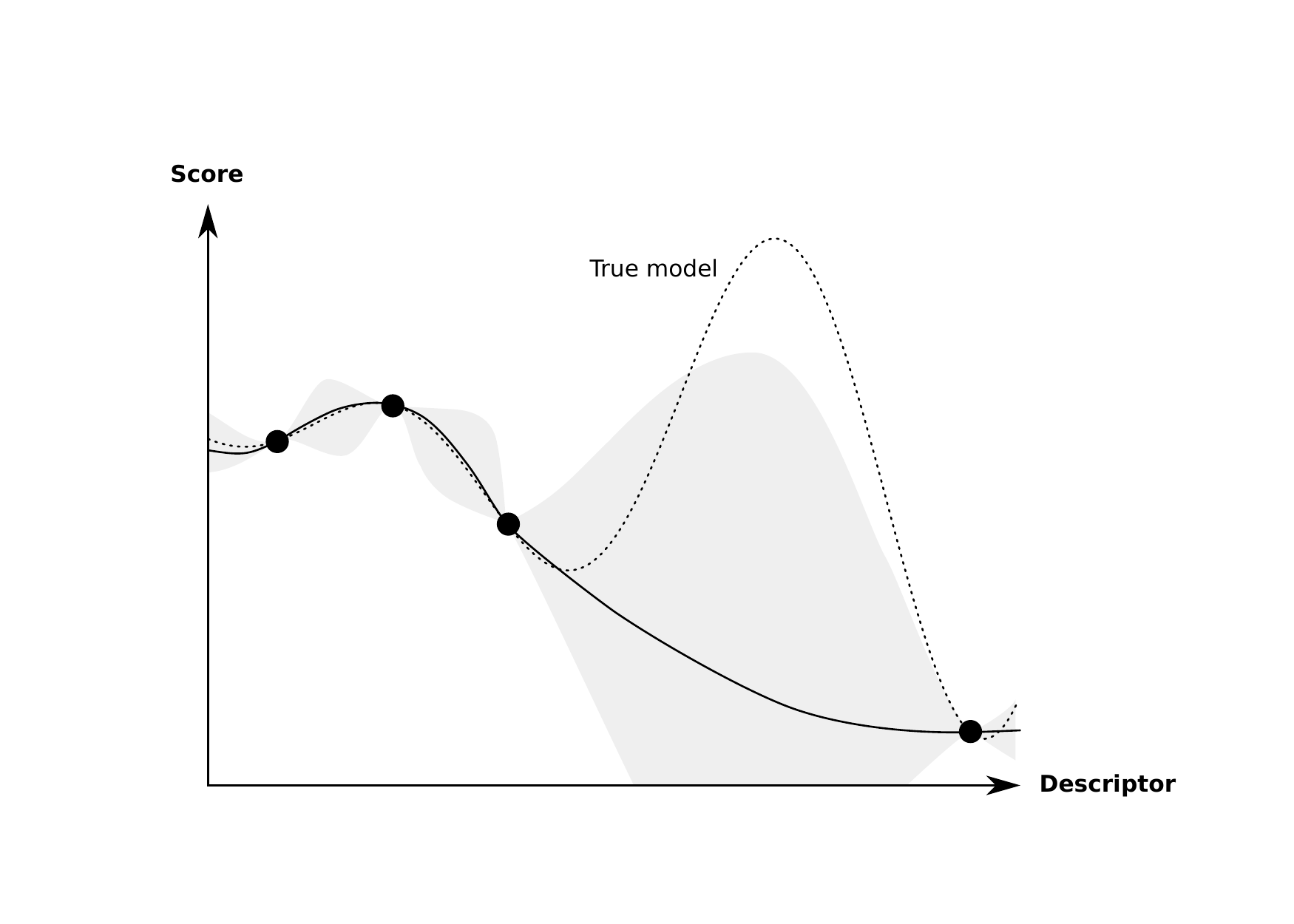}
\caption{Schematic illustration of a typical situation
in which Bayesian optimization works efficiently.
The dashed line represents the true model; the black circles denote
sampling points; the solid line shows a model obtained from 
the sampling points, and the gray region shows a confidence interval
of the model.}
\label{BO_illustration}
\end{figure}

 Bayesian optimization (BO) is a powerful technique to find the maximum
 (or the minimum) of an unknown function along this idea.
 It is based on 
 Bayesian modeling using a dataset collected in
 the previous sampling--modeling iterations,
 and does not require explicit form of the function $y=f(x)$.
 This method is efficient because it takes account of 
 the uncertainty of a model 
 in addition to the mean value.
 Figure \ref{BO_illustration} illustrates a typical situation in which BO is efficient. 
 The dashed line is the true model. 
 Suppose we have four sampled points which are denoted by black circles.
 By Bayesian modeling, we obtain the mean value (solid line) and the uncertainty (gray region). 
 In this situation, the mean value does not give good prediction for the highest-score point. 
 However, by considering the information of the uncertainty,
 one can find a significant probability that the true model
 has the maximum between the two rightmost data points.

 BO has been recently applied in various problems
 in materials science.\cite{Ueno16,Ju17,Kikuchi18}
 It has also a potential for application to
 optimization of a chemical composition,\cite{Yuan18}
 but there was no reports on
 quantitative estimation of efficiency
 in such a problem
 avoiding possible overestimation by mere luck
 to our knowledge.
 In such applications,
 we need to properly choose a search space,
 a descriptor for the candidate systems, 
 and a score to describe the performance
 that are suitable for the problem.
 Otherwise, the efficiency of the scheme is deteriorated.
 A descriptor---a form to which the input data is encoded---is
 especially crucial.\cite{Oganov09,Ghiringhelli15,Pham17,Pham18,Dam18,Ouyang18} 

 Accuracy of the first-principles calculation
 is also of great importance
 in the computer-aided materials search. 
 However,
 conventional methods are often insufficient to achieve enough accuracy
 while sophisticated schemes are too much time-consuming for the purpose.
 For example,
 the magnetic transition temperature is overestimated in the mean-field approximation. 
 Systematic errors also come in from numerical factors, such as a
 limited number of basis functions.

 It is one of promising ideas
 to improve the data
 by using a smaller dataset from more accurate
 calculations or experiments.\cite{Kennedy00,Pilania17}
 This idea is also seen in the notion of
 transfer learning, which uses referential datasets that are different
 from the target dataset,
 and transfer the knowledge from
 the reference to the target.\cite{Pan10,Hutchinson17}
 However, there is no method that works for any purposes,
 and we need to devise a method that is suitable for each of
 the problems on the basis of knowledge about the origin
 of the error.

 In this paper,
 we propose a practical framework for optimizing nonstoichiometric composition of multinary compounds
 based on Bayesian optimization and first-principles calculation.
 We perform a benchmark of our scheme and discuss
 its efficiency in the optimization
 using a dataset  
 obtained by first-principles calculation with the KKR-CPA method
 for systems with nonstoichiometric compositions.
 We investigate the performance of descriptors,
 and discuss how we can choose an efficient one for problems.
 To set up a pragmatic problem, we deal with an issue from the 
 cutting-edge of the materials study on hard magnets.
 We also present a method for correcting systematic errors
 in the formation energy by using
 smaller but more accurate dataset.
 Our idea is to construct a model of errors
 on the basis of our understanding of it.

 This paper is organized as follows: in section \ref{Problem_Setup},
 we describe our problem setup for the benchmark, providing 
 the background. In the first part of Section \ref{Methods},
 we present a brief summary of the whole framework.
 We then provide details of Bayesian optimization and
 the first-principles calculation
 in the subsequent subsections.
 Section \ref{Suriawase} is devoted to the method for
 integrating datasets that we use to improve the formation energy.
 We present the results of the benchmark and the data integration
 in Section \ref{Results}.
 Finally, we conclude the paper with a summary in Section \ref{Conclusion}.
 
\section{Problem setup and its background}
\label {Problem_Setup}

 $R$Fe$_{12}$-type compounds having the ThMn$_{12}$ structure
 have been considered
 as a possible main phase of a strong hard-magnet
 because they are expected to have high magnetization due to
 its high Fe content, and 
 to have high magnetocrystalline anisotropy
 if the $R$ element is properly
 chosen.\cite{Ohashi88,Ohashi88b,Yang88,Verhoef88,
 DeMooij88,Buschow88,Jaswal90,Coehoorn90,
 Buschow91,
 Sakurada92,Sakuma92,
 Asano93,Akayama94,
 Kuzmin99,
 Gabay16,
 Koener16,Ke16,Fukazawa17,Fukazawa19}
 The magnetic properties of NdFe$_{12}$N was evaluated 
 theoretically a few years ago \cite{Miyake14}, and its high magnetization and
 anisotropy field were confirmed by a successful synthesis
 of NdFe$_{12}$N$_x$ film.\cite{Hirayama15,Hirayama15b}

 Unfortunately, NdFe$_{12}$N and its mother compound NdFe$_{12}$ are thermally unstable. 
 They cannot be synthesized as a bulk without substituting another element for
 a part of the Fe elements.
 Titanium is typical of such a stabilizing element.\cite{Yang91,Yang91b}
 Introduction of Ti, however, reduces the magnetization significantly. 
 Co also has a potential to work as
 a stabilizing element according to a prediction by
 first-principles calculation.\cite{Harashima16}
 Compared to Ti, Co is favorable in terms of magnetization.
 In fact,
 a recent experiment on  
 Sm(Fe$_{0.8}$Co$_{0.2}$)$_{12}$ film showed that it has superior saturation
 magnetization and anisotropy field to Nd$_2$Fe$_{14}$B, the main phase of
 the current strongest magnet.\cite{Hirayama17}
 Chemical substitution at the $R$ site also affects structural stability. 
 Zirconium has been attracted attention as a stabilizing element at the rare-earth site.\cite{Suzuki14,Sakuma16,Kuno16,Suzuki16}
 Recent first-principles calculation predicted that Dy also works as a stabilizer.\cite{Harashima18}
 Therefore, optimization of chemical composition of $R$Fe$_{12}$-type compounds in terms of stability and magnetic properties is an important issue for the development of next-generation permanent magnets.

 Bearing these in mind, we set 
 $R$Fe$_{12}$-type magnet compounds
 as target systems.
 Especially speaking,
 we optimize chemical formula of
  (R$_{1-\alpha}$Z$_{\alpha}$)(Fe$_{1-\beta}$Co$_{\beta}$)$_{12-\gamma}$Ti$_{\gamma}$
 (R = Y, Nd, Sm; Z = Zr, Dy) 
 so that it maximizes magnetization, the Curie temperature,
 or minimizes the formation energy from the unary systems
 in the benchmark.
 Therefore, the problem is a combination of optimization with respect to
 the compositions ($\alpha$, $\beta$ and $\gamma$) and 
 optimization with respect to
 the choice of elements for R and Z.
 We discuss the efficiency of the optimization by comparing
 a number of iterations required
 in Bayesian optimization with that in random sampling. 
 We also study how the efficiency is affected by the choice of descriptor. 
 
\section{Methods}
\label{Methods}
\begin{figure}[htbp]
\centering
\includegraphics[bb=0 0 504 360, width=9cm]{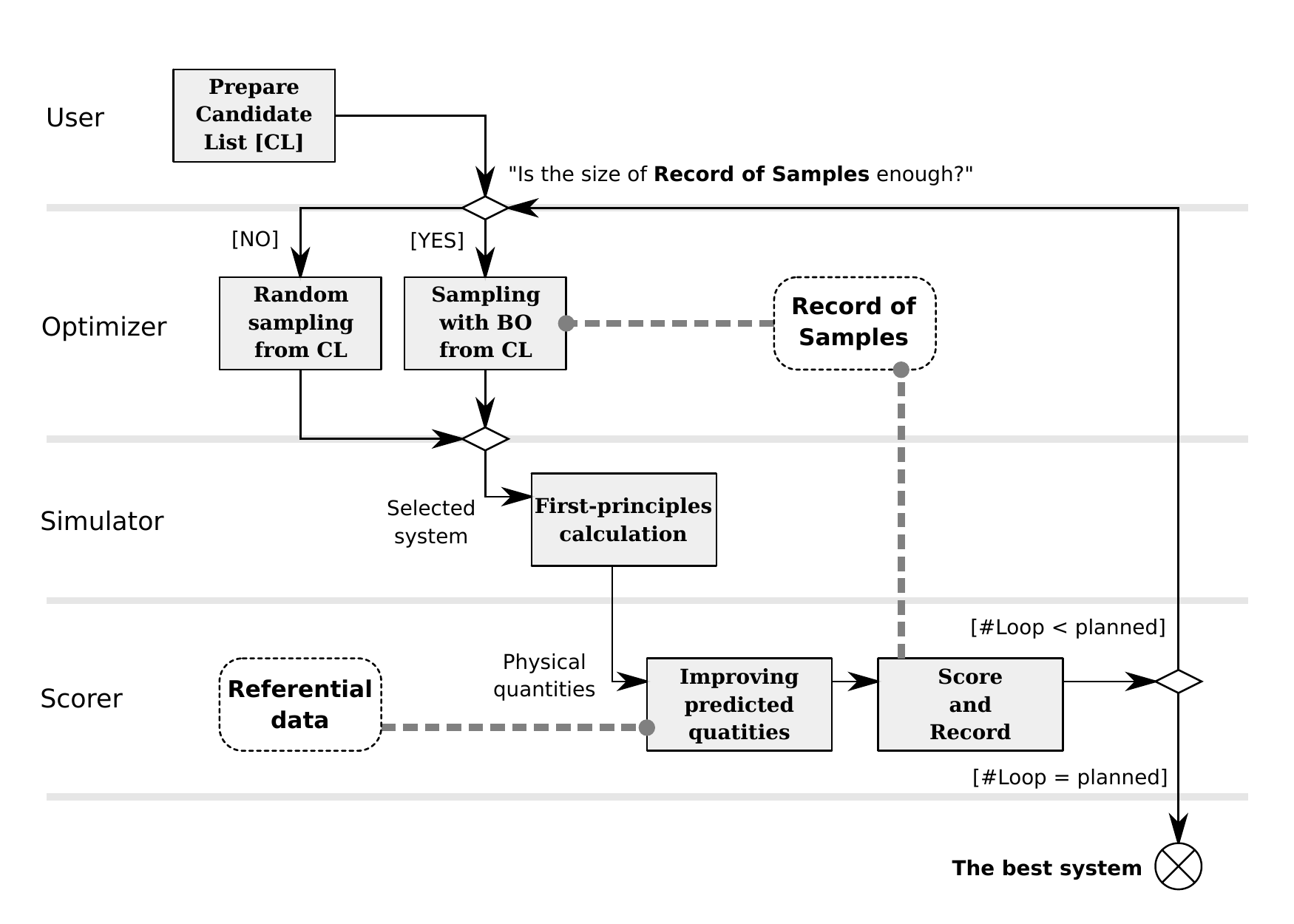}
\caption{The workflow of the proposed scheme for materials search.
The solid squares denote activities; the dashed squares with round corners
represent datasets. BO stands for Bayesian optimization; CL stands for 
candidate list.}
\label{Scheme_Diagram}
\end{figure}
Figure~\ref{Scheme_Diagram} shows the workflow in our optimization framework.
At the beginning of the scheme,
the user
prepares a list of candidate compounds. 
The candidates are expressed 
in the form of a descriptor.
In this study,
we prepare 11 types of descriptors
for the
(R$_{1-\alpha}$Z$_{\alpha}$)(Fe$_{1-\beta}$Co$_{\beta}$)$_{12-\gamma}$Ti$_{\gamma}$ (R=Y, Nd, Sm; Z=Zr, Dy) 
systems,
which we 
discuss in Section \ref{Sec_BO}.

Then, the candidate list is passed to the optimizer.
The role of the optimizer is to pick one system from the candidate list
so that a system with a high score is quickly found
in the whole scheme.
Because it does not have enough data
to perform Bayesian optimization at the beginning,
it randomly chooses a system from the list.
It receives a feedback from a scorer later in the scheme,
and record it.
When the record reaches a certain size, 
the sampling method is switched to Bayesian optimization.
To cover the role of the optimizer, we use a Python module called
``COMmon Bayesian Optimization library'' (COMBO).\cite{Ueno16,COMBO}
We present the parameters used in our benchmark
in Section \ref{Sec_BO}.

In the next stage, a quantum simulator calculates physical properties
for the system chosen, which is the most time-demanding process in the scheme.
The details in our simulation is described in
Section \ref{Sec_FP}.

Then, the scorer integrates the calculated properties to a score.
It also performs improvement of the estimated values
by using the referential data
before generating the score.
In our application,
we use the value of magnetization, Curie temperature or 
the formation energy as a score.
As for the formation energy,
we improve it
by the method for integrating datasets
presented in Section \ref{Suriawase}.
The preparation of the referential data is described in Section \ref{Sec_FP}.
The score is fed back to the optimizer to increase the size of the data
used in Bayesian optimization.

The iteration loop is repeated
until
the number of iterations reaches a criterion. 
Otherwise, the workflow goes back to the optimizer. 
After the loop has ended,
the candidate with the best score found in the iterations
is output.

\subsection{Bayesian optimization}
\label{Sec_BO}
As mentioned above, we use COMBO\cite{Ueno16,COMBO}
to cover the role of 
the optimizer in Fig.~\ref{Scheme_Diagram}.
We use Thompson sampling as a heuristic
to the exploration--exploitation problem
in optimization.
The dimension of the random feature maps,
which determines the degree of approximation for 
the Gaussian kernel, is set to 5000.
The first 10 samples are chosen randomly
without using Bayesian optimization.
The number of iterations is set to 100, including the 
first 10 iterations with the random sampling.

The candidate list consists of
 (R$_{1-\alpha}$Z$_{\alpha}$)(Fe$_{1-\beta}$Co$_{\beta}$)$_{12-\gamma}$Ti$_{\gamma}$ systems
for all the combination of R=Y, Nd, Sm; Z=Zr, Dy;
$\alpha=0, 0.1, \cdots, 1$; $\beta=0, 0.1, \cdots, 1$; $\gamma=0, 0.5,
\cdots, 2$.
There are duplication in the list
[e.g. YZr$_0$Fe$_{12}$ and YDy$_0$Fe$_{12}$],
and 
the number of the unique items is 3245
out of the $3\times 2 \times 11 \times 11 \times 5 = 3630$ systems.

We use 11 different sets of descriptors listed in Table \ref{tab_descriptors}.
The descriptors consist of the number of electrons per cell ($N$),
the number of electrons from the R element per cell ($N_\mathrm{R}$),
the number of electrons from the Z element per cell ($N_\mathrm{Z}$),
$N_\mathrm{R} + N_\mathrm{Z}$ ($\equiv N_\mathrm{2a}$),
the number of electrons from the transition elements---namely Fe,Co,Ti---per cell ($N_\mathrm{T}$),
the atomic number of the R element ($Z_\mathrm{R}$),
the atomic number of the Z element ($Z_\mathrm{Z}$),
an index for the R element ($n_\mathrm{R}$ = 0, 1, 2 corresponding to R
= Y, Nd, Sm),
an index for the Z element ($n_\mathrm{Z}$ = 0, 1    corresponding to Z
= Zr, Dy),
the Z content per cell ($\alpha$),
the Co/(Fe+Co) ratio ($\beta$),
the Ti content ($\gamma$),
and the values of 
$\alpha_1$, $\alpha_2$, $\alpha_3$, and $\alpha_4$
when the chemical formula is expressed in the form of 
 (Y$_{1-\alpha_1-\alpha_2-\alpha_3-\alpha_4}$Nd$_{\alpha_1}$Sm$_{\alpha_2}$Zr$_{\alpha_3}$Dy$_{\alpha_4}$)
(Fe$_{1-\beta}$Co$_{\beta}$)$_{12-\gamma}$Ti$_{\gamma}$.
\begin{table}[htbp]
\caption{11 Descriptors used in the Bayesian optimization. See the text 
for description of the variables, $N$, $N_\mathrm{R}$, $N_\mathrm{Z}$, $N_\mathrm{2a}$, $N_\mathrm{T}$,
$Z_\mathrm{R}$, $Z_\mathrm{Z}$, $n_\mathrm{R}$, $n_\mathrm{Z}$,
$\alpha$, $\beta$, $\gamma$, $\alpha_1$, $\alpha_2$, $\alpha_3$ and $\alpha_4$.
\label{tab_descriptors}}
\begin{tabular}{cc|cc}
 \hline
 \hline
 \#1 & $N$& --- & --- \\ \#2   & $N_\mathrm{2a}$, $N_\mathrm{T}$&
   \#7 & $N_\mathrm{2a}$, $\beta$, $\gamma$\\
 \#3   & $N_\mathrm{R}$,  $N_\mathrm{Z}$, $N_\mathrm{T}$&
   \#8 & $N_\mathrm{R}$,  $N_\mathrm{Z}$, $\beta$, $\gamma$\\
 \#4   & $Z_\mathrm{R}$,  $Z_\mathrm{Z}$, $\alpha$, $N_\mathrm{T}$ &
   \#9 & $Z_\mathrm{R}$,  $Z_\mathrm{Z}$, $\alpha$, $\beta$, $\gamma$ \\
 \#5   & $n_\mathrm{R}$,  $n_\mathrm{Z}$, $\alpha$, $N_\mathrm{T}$ &
   \#10& $n_\mathrm{R}$,  $n_\mathrm{Z}$, $\alpha$, $\beta$, $\gamma$ \\
 \#6   & $\alpha_1$, $\alpha_2$, $\alpha_3$, $\alpha_4$, $N_\mathrm{T}$ &
   \#11& $\alpha_1$, $\alpha_2$, $\alpha_3$, $\alpha_4$, $\beta$, $\gamma$ \\
 \hline
 \hline
\end{tabular}
\end{table}

\subsection{First-principles calculation}
\label{Sec_FP}
In the ``Simulator'' block 
in Fig.\ref{Scheme_Diagram},
we perform first-principles calculation
based on
density functional theory
with the local density approximation.\cite{Hohenberg64,Kohn65}
We use
the open-core approximation\cite{Jensen91,Richter98,Locht16}
to the f-electrons in Nd, Sm and Dy
and apply the self-interaction correction.\cite{Perdew81}

\begin{figure}[htbp]
\centering
\includegraphics[bb=0 0 1200 1200, width=7.5cm]{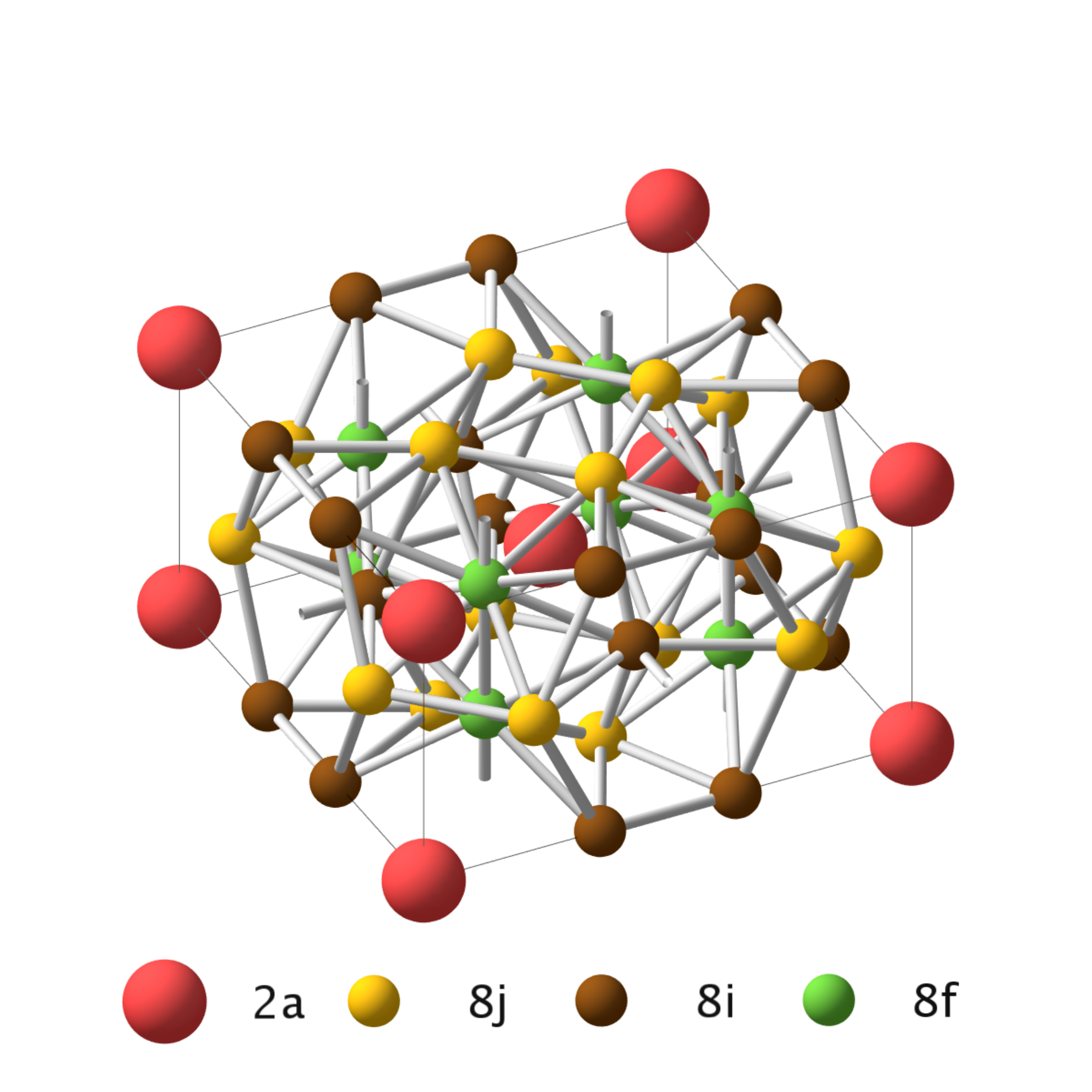}
\caption{The 2a, 8j, 8i, and 8f Wycoff positions in the ThMn$_{12}$ structure.
 Some bonds are shown as eye-guides to see the three-dimensional structure.}
\label{fig_structure}
\end{figure}

We assume the ThMn$_{12}$ structure (Fig.~\ref{fig_structure}) for the
(R$_{1-\alpha}$Z$_{\alpha}$)(Fe$_{1-\beta}$Co$_{\beta}$)$_{12-\gamma}$Ti$_{\gamma}$
systems. 
The lattice parameters 
are determined by linear 
interpolation from 
those for 
RFe$_{12}$, RFe$_{11}$Ti, 
ZFe$_{12}$, ZFe$_{11}$Ti and
RCo$_{12}$.
These values for the stoichiometric systems 
were calculate with
the PAW method\cite{Bloechl94,Kresse99}
using a software package QMAS\cite{QMAS}.
We use 
the PBE exchange-correlation functional\cite{Perdew96}
of the Generalized
Gradient Approximation (GGA)
to obtain adequate structures.
The values for ZFe$_{11}$Ti and RCo$_{12}$ are presented in Appendix
\ref{lattparams}.
Those for RFe$_{12}$, RFe$_{11}$Ti and ZFe$_{12}$ are in Refs.\cite{Fukazawa18,Harashima14b,Harashima18}.

In the treatment with the coherent potential approximation (CPA)\cite{Soven67,Soven70,Shiba71},
we assume quenched randomness for random occupation of the elements:
the element R and Z are assumed to occupy the 2a site
(see Fig.~\ref{fig_structure} for the Wycoff positions).
Titanium is assumed to occupy the 8i site.
Iron (cobalt) is assumed to occupy the 8f, 8i and 8j site with a common
probability of $1-\beta$ ($\beta$) to these sites.

We calculate the magnetization,
the Curie temperature, and the formation energy from the 
unary systems. 
We use the raw value of magnetization from KKR-CPA.
To estimate the Curie temperature, 
we calculate intersite magnetic couplings
by using Liechtenstein's method,\cite{Liechtenstein87}
and convert them to the Curie temperature
using the mean-field approximation.\cite{Fukazawa18}
Although this procedure overestimates the Curie temperature,
we can expect from previous results
that it can capture materials trend
because theoretical values within
the mean-field approximation
had a significant correlation
with experimental Curie temperatures.\cite{Fukazawa19b}

The best value among the candidates 
is 1.76 T [DyFe$_{12}$] for magnetization,
1310 K [Sm(Fe$_{0.2}$Co$_{0.8}$)$_{12}$]
for Curie temperature,
and $-2.85$ eV [SmCo$_{10}$Ti$_2$]
for the formation energy.
It should be noted, however, that
the values on the list cannot be directly used  
as information for experimental synthesis
because the data do not include information
of phase competition (especially with
Th$_2$Zn$_{17}$-type
and Th$_2$Ni$_{17}$-type
phases),
magnetic anisotropy,
and contribution to the magnetization 
from the f-electrons.
We cover this subject in Appendix \ref{rel2exp},
and provide lists of some of the best systems
with the physical properties there.

As for the formation energy, 
KKR
needs too large computational resources
to obtain an accurate energy difference between systems
when they have far different structures from each other.
We use the method that
we describe in the following subsection
to correct the energy difference calculated by KKR-CPA
with referential data of total energy obtained by PAW.

\subsection{A method for integration datasets}
\label{Suriawase}
Let us consider
the formation energy from the unary systems defined as follows:
\begin{align}
 \Delta E
  &\equiv E[\text{(R$_{1-\alpha}$Z$_{\alpha}$)(Fe$_{1-\beta}$Co$_{\beta}$)$_{12-\gamma}$Ti$_{\gamma}$}]
 \nonumber \\
 & - E[\text{(the unary systems)}],
 \label{Def_Formation}
\end{align}
where ``(the unary systems)'' is defined as
\begin{align}
 &(\text{the unary systems})
 \nonumber \\
  &=
  (1-\alpha) \text{R}
  + \alpha \text{Z}
 \nonumber \\
  &\quad + (1-\beta)(12-\gamma) \text{Fe}
  + \beta(12-\gamma)     \text{Co}
  + \gamma \text{Ti},
\end{align}
and $E[\cdot]$ denotes the total energy of the system in the square bracket.
Because
the structures of
(R$_{1-\alpha}$Z$_{\alpha}$)(Fe$_{1-\beta}$Co$_{\beta}$)$_{12-\gamma}$Ti$_{\gamma}$
and each of the unary systems are
much different from one another, 
it is not efficient to directly calculate this formation energy
with the KKR method, 
although it can deal with non-stoichiometric systems with CPA.
Our idea is to calculate the formation energy of stoichiometric systems 
more accurately by another method, and use calculated energies   
as reference data.

We construct a stochastic model
for the total energy of
(R$_{1-\alpha}$Z$_{\alpha}$)(Fe$_{1-\beta}$Co$_{\beta}$)$_{12-\gamma}$Ti$_{\gamma}$
based on the expectation 
that the smaller structural difference two systems have,
the more accurate energy difference KKR-CPA gives.
To quantify the difference of systems,
we consider a descriptor
with which
the difference between the systems ($\vec{x}$ and $\vec{y}$)
is well-described by the distance ($|\vec{x}-\vec{y}|$).
Let us denote the reference systems
in the form of the descriptor
by $\vec{x}^\text{\,R}_1, \vec{x}^\text{\,R}_2, \cdots, \vec{x}^\text{\,R}_M$
where $M$ is the number of the reference systems.
The descriptor here does not have to be identical to the descriptor 
used in the Bayesian optimization.
In the demonstration, we use a set of
$(\alpha',\beta',\gamma')\equiv
(\alpha, \beta(12-\gamma)/12, \gamma/12)$
with which the search space can be expressed as
(R$_{1-\alpha'}$Z$_{\alpha'}$)(Fe$_{1-\beta'-\gamma'}$Co$_{\beta'}$Ti$_{\gamma'}$)$_{12}$
irrespective of the choice of the descriptor in the optimization.

\begin{figure}[htbp]
\centering
\includegraphics[bb=0 0 504 360, width=9cm]{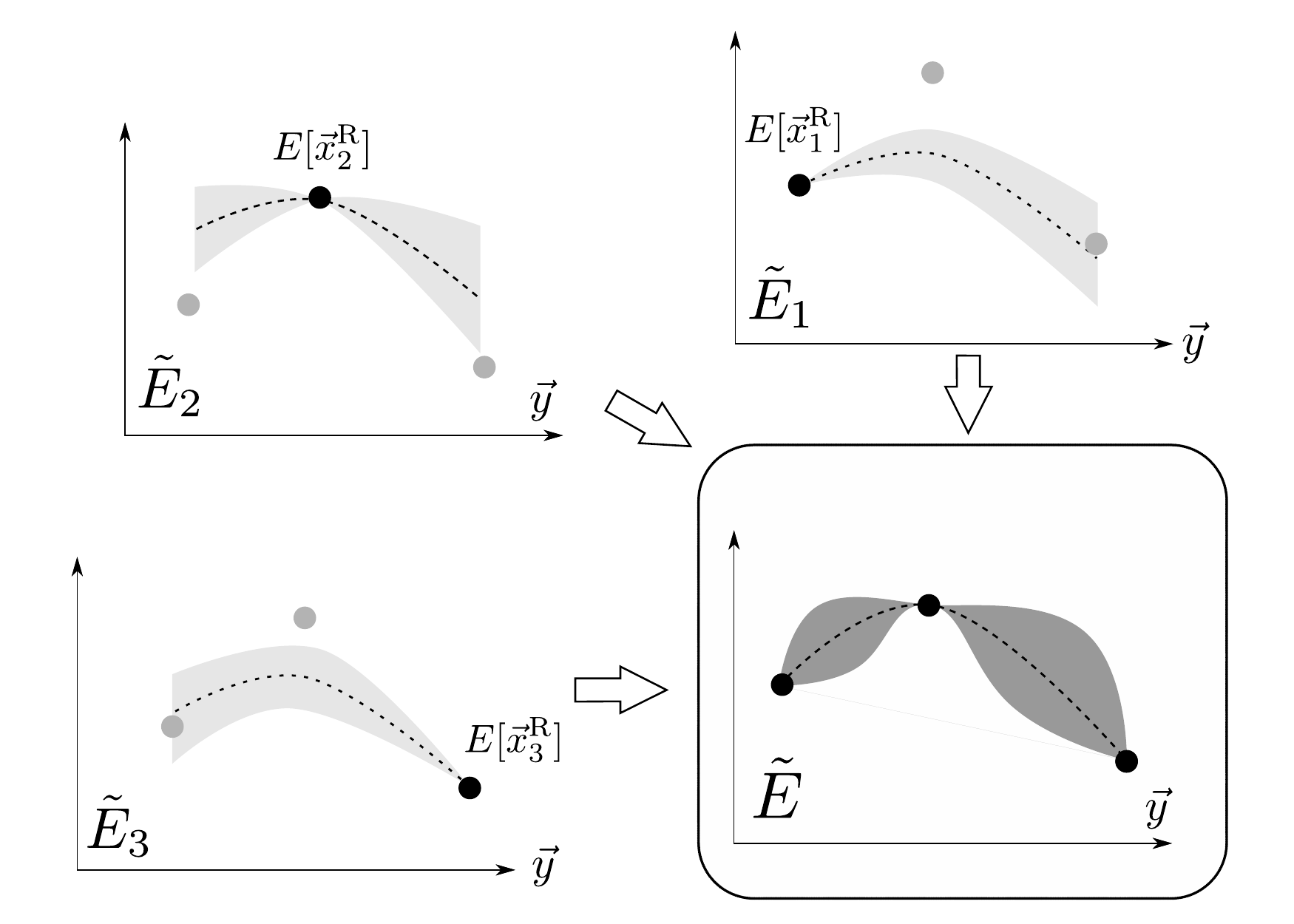}
\caption{A schematic diagram of the procedure for constructing
an integrated model. The three graphs outside the box denote
the models defined by Eq.~\eqref{model_atomic}.
They are integrated into the model described
by Eq.~\eqref{model_integrated}.}
\label{suriawase}
\end{figure}

For each of the reference points $\vec{x}^\text{\,R}_i$,
we construct a stochastic model $\tilde{E}_i[\vec{y}]$
for the total energy of a system $\vec{y}$
(see also the graphs outside the box in Fig.~\ref{suriawase}):
\begin{equation}
 \tilde{E}_i[\vec{y}]
  -
  E[\vec{x}^\text{\,R}_i]
  =
  E'[\vec{y}]
  -
  E'[\vec{x}^\text{\,R}_i]
  +
  \varepsilon_i.
  \label{model_atomic}
\end{equation}
The two $E$'s in the right-hand side (to which primes are attached)
are the total energy calculated with KKR-CPA, whereas
$E[\vec{x}^\text{\,R}_i]$ in the left-hand side
is evaluated by a more accurate method, 
for which we use PAW in the present work.
$\varepsilon_i$ is a random variable whose distribution is 
$\mathcal{N}(0,S^2_i)$, i.e.
the normal distribution whose mean is zero and variance is $S^2_i$,
where 
$S^2_i\equiv\sigma^2\left|\vec{y} - \vec{x}^\text{\,R}_i\right|$ and 
$\sigma^2$ is a parameter we will estimate later.
This model describes the expectation that 
the deviation of the energy difference
(the first two terms in the right-hand side)
from the true difference
(the left-hand side)
tends to be large when the
difference,
$\left| \vec{y} - \vec{x}^\text{\,R}_i \right|$,
is large.
The graphs outside the box
in Fig.~\ref{suriawase} depict how $\tilde{E}_i$
behave: 
there are three models corresponding to the three
reference systems
($\vec{x}^\mathrm{R}_1$ ,$\vec{x}^\mathrm{R}_2$, $\vec{x}^\mathrm{R}_3$),
and 
the error $\varepsilon_i$ in each model is large 
when the distance of the reference point from
$\vec{y}$ is large.

We then integrate these models,
$\tilde{E}_1, \cdots, \tilde{E}_M$,
into a single model $\tilde{E}$
by imposing the following condition to
the distribution of $\varepsilon_i$:
\begin{equation}
 \tilde{E}_1[\vec{y}]
  =
 \tilde{E}_2[\vec{y}]
  =
  \cdots
  =
 \tilde{E}_M[\vec{y}]
  \equiv
 \tilde{E}[\vec{y}].
\end{equation}
This condition can be rewritten as follows:
\begin{equation}
 \varepsilon_i
  =
 \tilde{E}[\vec{y}]
 -
 E[\vec{x}^\text{\,R}_i]
 -
  E'[\vec{y}]
  +
  E'[\vec{x}^\text{\,R}_i].
\end{equation}
Therefore, the conditional probability distribution of $\tilde{E}$ is
\begin{equation}
 \Pr \left( \tilde{E}[\vec{y}] = t \right)
  =
  \frac{
  \prod_i
  P_i
  \left(
      t
      -
      E[\vec{x}^\text{\,R}_i]
      -
      E'[\vec{y}]
      +
      E'[\vec{x}^\text{\,R}_i]
  \right)
  }
  {
  \int
  dt'
  \prod_i
  P_i
  \left(
      t'
      -
      E[\vec{x}^\text{\,R}_i]
      -
      E'[\vec{y}]
      +
      E'[\vec{x}^\text{\,R}_i]
  \right)
  }
\end{equation}
where $P_i$ denotes the probability distribution function of 
$\mathcal{N}(0,S^2_i)$.
It is straightforward to see that 
this conditional distribution is the normal distribution
whose mean $\tilde{\mu}$ and variance $\tilde{S}^2$ are
\begin{equation}
 \tilde{\mu}
  =
  E'[\vec{y}]
  +
  \frac{1}{\Omega}
  \sum_i
  \omega_i
  \left\{
      E[\vec{x}^\text{\,R}_i]
      -
      E'[\vec{x}^\text{\,R}_i]
  \right\},
  \label{Suriawase_mu}
\end{equation}
\begin{equation}
 \tilde{S}^2
  =
  \frac{\sigma^2}{\Omega}
  \label{Suriawase_S2}
\end{equation}
where $\omega_i$ is a weight and $\Omega$ is a normalization factor
that are defined as 
\begin{equation}
 \omega_i
  \equiv
  \frac{1}{\left| \vec{y} - \vec{x}^\text{\,R}_i \right|},
 \quad
 \Omega \equiv \sum_i \omega_i.
\label{normalization}
\end{equation}
In another expression, our integrated model is
\begin{equation}
 \tilde{E}[\vec{y}]
  =
  E'[\vec{y}]
  +
  \frac{1}{\Omega}
  \sum_i
  \omega_i
  \left\{
      E[\vec{x}^\text{\,R}_i]
      -
      E'[\vec{x}^\text{\,R}_i]
  \right\}
  +
  \tilde{\varepsilon}
\label{model_integrated}
\end{equation}
where $\tilde{\varepsilon}$ is a random variable whose distribution is
$\mathcal{N}(0,\tilde{S}^2)$.
Algorithm \ref{Algorithm_1} summarizes the construction of the model
explained so far
in a form of a pseudocode.
Note that the value for the input variable $\sigma^2$ has not yet been determined.
The characteristics of $\tilde{E}[\vec{y}]$ is illustrated in the right-bottom panel 
in Fig.~\ref{suriawase}. Although $\tilde{E}[\vec{y}]$ is singular
at $\vec{y} = \vec{x}_i^\mathrm{R}$, 
it is easy to see that this is removable and
$\lim_{\vec{y}\rightarrow \vec{x}_i^\mathrm{R}}\tilde{E}[\vec{y}] = E[\vec{x}_i^\mathrm{R}]$.
\renewcommand{\algorithmicrequire}{\textbf{Input:}}
\renewcommand{\algorithmicensure}{\textbf{Output:}}
\begin{algorithm}[H]
\begin{algorithmic}[]
\REQUIRE
 $\vec{x}^\mathrm{R}_i$ ($i = 1, \cdots, M$),
 $\vec{y} \not\in \{\vec{x}^\mathrm{R}_i\}$ :  descriptor;
 $\sigma^2$;\\
 $E'[\vec{y}]$: energy estimation;
 $\Delta_i \equiv E[\vec{x}^\mathrm{R}_i] - E'[\vec{x}^\mathrm{R}_i]$
\ENSURE $\tilde{\mu}$, $\tilde{S}^2$
\STATE Initialize $\tilde{\mu} \leftarrow 0$, $\Omega \leftarrow 0$
    \FOR{$i = 1$ to $M$}
        \STATE $\Omega \leftarrow \Omega + 1/|\vec{y} - \vec{x^\mathrm{R}_i}|$
        \STATE{$\tilde{\mu} \leftarrow \tilde{\mu} + \Delta_i / |\vec{y} - \vec{x^\mathrm{R}_i}|$}
    \ENDFOR
    \STATE {\it Average}: $\tilde{\mu} \leftarrow \tilde{\mu} / \Omega + E'[\vec{y}]$ \, [Eq.~\eqref{Suriawase_mu}]
    \STATE {\it Variance}: $\tilde{S}^2 \leftarrow \sigma^2 / \Omega$ \, [Eq.~\eqref{Suriawase_S2}]
\end{algorithmic}
\caption{\label{Algorithm_1}
A pseudocode of the algorithm for determination of 
$\tilde{\mu}$ in Eq.~~\eqref{Suriawase_mu}
and $\tilde{S}^2$ in Eq.~\eqref{Suriawase_S2}.
This needs $\sigma^2$ and $\Delta_i$ as inputs, which are 
calculated by the algorithm shown in Alg.~\ref{Algorithm_2}.}
\end{algorithm}
The formation energy from the unary systems can be calculated as
$\Delta E \simeq \tilde{E}[\vec{y}] - E[(\text{the unary systems})]$, 
where $E[(\text{the unary systems})]$ is calculated by an accurate method.

To complete the formulation, we discuss estimation of $\sigma^2$
based on the data for the reference systems.
We use the maximum likelihood estimation.
However, it cannot be directly applied to our model 
because the distribution of $\tilde{\varepsilon}$ becomes
the delta function
in the limit of $\vec{y} \rightarrow \vec{x}^\text{\,R}_i$,
regardless of the value of $\sigma^2$.
\begin{figure}[htbp]
\centering
\includegraphics[bb=0 0 504 360, width=7cm]{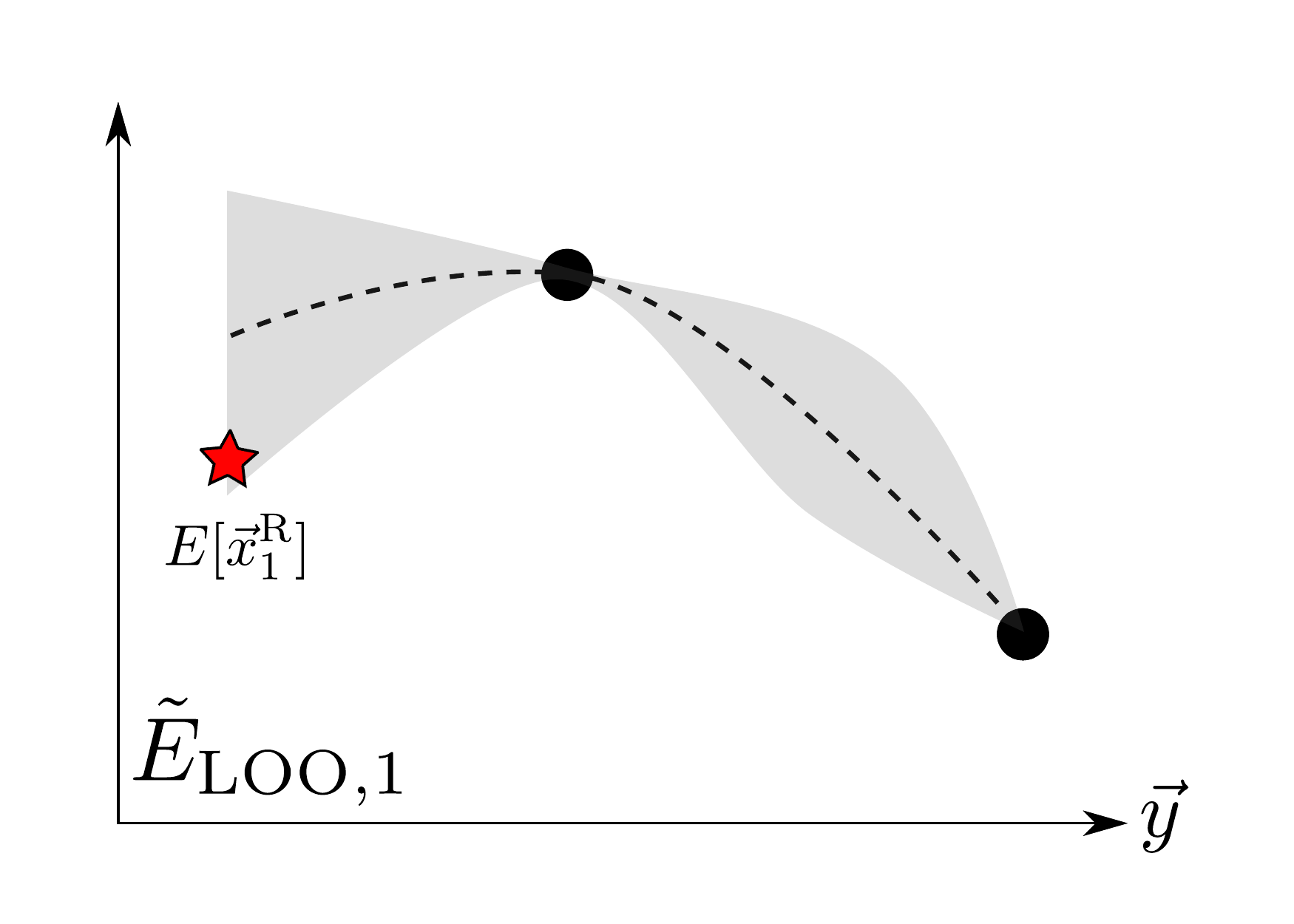}
\caption{A schematic plot of $\tilde{E}_{\text{LOO},i}$
for $i=1$ and $M=3$.
}
\label{suriawase2}
\end{figure}
To avoid this singularity, we consider a model
$\tilde{E}_{\text{LOO},i}$ that is constructed 
from all reference systems but $\vec{x}^\text{\,R}_i$.
Figure \ref{suriawase2} depicts the construction of such a model.
Now, we consider the probability of $\tilde{E}_{\text{LOO},i}[\vec{y}]$ 
at $\vec{y} = \vec{x}^\text{\,R}_i$.
Regarding 
the probability for 
$\tilde{E}_{\text{LOO},i}[\vec{y}= \vec{x}^\text{\,R}_i] = E[\vec{x}^\text{\,R}_i]$
as a likelihood $L_i$, 
we select the value of $\sigma^2$
that maximizes $\mathcal{L}=\prod_i L_i$.
We then obtain 
\begin{equation}
 \sigma^2
  =
  \frac{1}{M}
  \sum_i
  \Omega_{\text{LOO},i}
  \left(
      E[\vec{x}^\text{\,R}_i]
      -
      E'[\vec{x}^\text{\,R}_i]
  \right)^2
\label{sigma}
\end{equation}
where 
\begin{equation}
 \Omega_{\text{LOO},i}
  \equiv
  \sum_{j\neq i}
  \frac{1}{\left| \vec{x}^\text{\,R}_i - \vec{x}^\text{\,R}_j \right|}.
\label{Omega_LOO}
\end{equation}
The determination of $\sigma^2$ is summarized 
in a form of a pseudocode in 
Alg.~\ref{Algorithm_2}.
The output values of $\sigma^2$ and $\Delta_i$ corresponds to
those in the algorithm 
described in Alg.~\ref{Algorithm_1}.
\begin{algorithm}[H]
\begin{algorithmic}[]
\REQUIRE
 $\vec{x}^\mathrm{R}_i$ ($i = 1, \cdots, M$):  descriptor;\\
 $E[\vec{x}^\mathrm{R}_i]$, $E'[\vec{x}^\mathrm{R}_i]$: energy estimation
\ENSURE $\Delta_i$($i = 1, \cdots, M$), $\sigma^2$
\STATE Initialize $\sigma^2 \leftarrow 0$ 
\FOR{$i = 1$ to $M$}
\STATE $\Delta_i \leftarrow E[\vec{x}^\mathrm{R}_i] - E'[\vec{x}^\mathrm{R}_i]$ 
\STATE $\sigma^2 \leftarrow \sigma^2 + \Delta_i^2 \cdot \sum_{j\neq i} 1/|\vec{x}^\mathrm{R}_i - \vec{x}^\mathrm{R}_j|$
\ENDFOR
\STATE $\sigma^2 \leftarrow \sigma^2 / M$ \, [Eq.~\eqref{sigma}]
\end{algorithmic}
\caption{\label{Algorithm_2}
A pseudocode of the algorithm for
determination of $\sigma^2$
in Eq.~\eqref{Suriawase_S2}.
The energy difference $\Delta_i$ is also output
to be used in the algorithm described in Alg.~\ref{Algorithm_1}.
}
\end{algorithm}

In the actual application,
we calculate
$E'[\vec{x}^\text{\,R}_i]$ with the KKR-LDA method
and $E[\vec{x}^\text{\,R}_i]$ with the PAW-GGA method.
We need to 
calibrate $E[\vec{x}^\text{\,R}_i]$ with a linear term
because
the difference in the treatment of the core electrons
is another major source of error in the total energy,
which is described well by a linear function.
We deal with it by extending 
our model to include an adjustable linear term,
and use it in the actual calculation.
We discuss this extension of the model
in Appendix \ref{Model_with_Linear_term}.

\section{Results and Discussion}
\label{Results}
\subsection{Integration method}
We show how
the integration model explained in Section \ref{Suriawase}
works before we present the benchmark of the whole scheme 
in the next subsection.
Let us take a simple example first: we consider a one dimensional
function  $E[x] = \sin(\pi x)$ as a true model.
Assume that we have many data about $E'[x]$, which is 
an approximate function and actually obey
$E'[x] = \sin(\pi x) + 2x$.
Because their derivatives differ from each other by 2,
the difference $E'[x] - E'[x^\text{R}_i]$ deviates
from $E[x] - E[x^\text{R}_i]$ roughly by 
$2(x-x^\text{R}_i)$.
The assumption of the model described in Section \ref{Suriawase}
was that the error in $E'[x]$ is large
when $|x-x^\text{R}_i|$ is large.
This is correct in the magnitude of the errors,
but there is a bias in its sign.
In this example, 
we prepare a dataset of $E'[x]$ for $x=-0.5$ to $0.5$
with an interval of 0.02.
We choose five reference points $x^\text{R}_i$ ($i=1,\cdots,5$)
from the $x$-values,
and prepare a dataset of $E[x^\text{R}_i]$.

\begin{figure}[htbp]
\includegraphics[width=7cm,bb = 0 0 504 360]{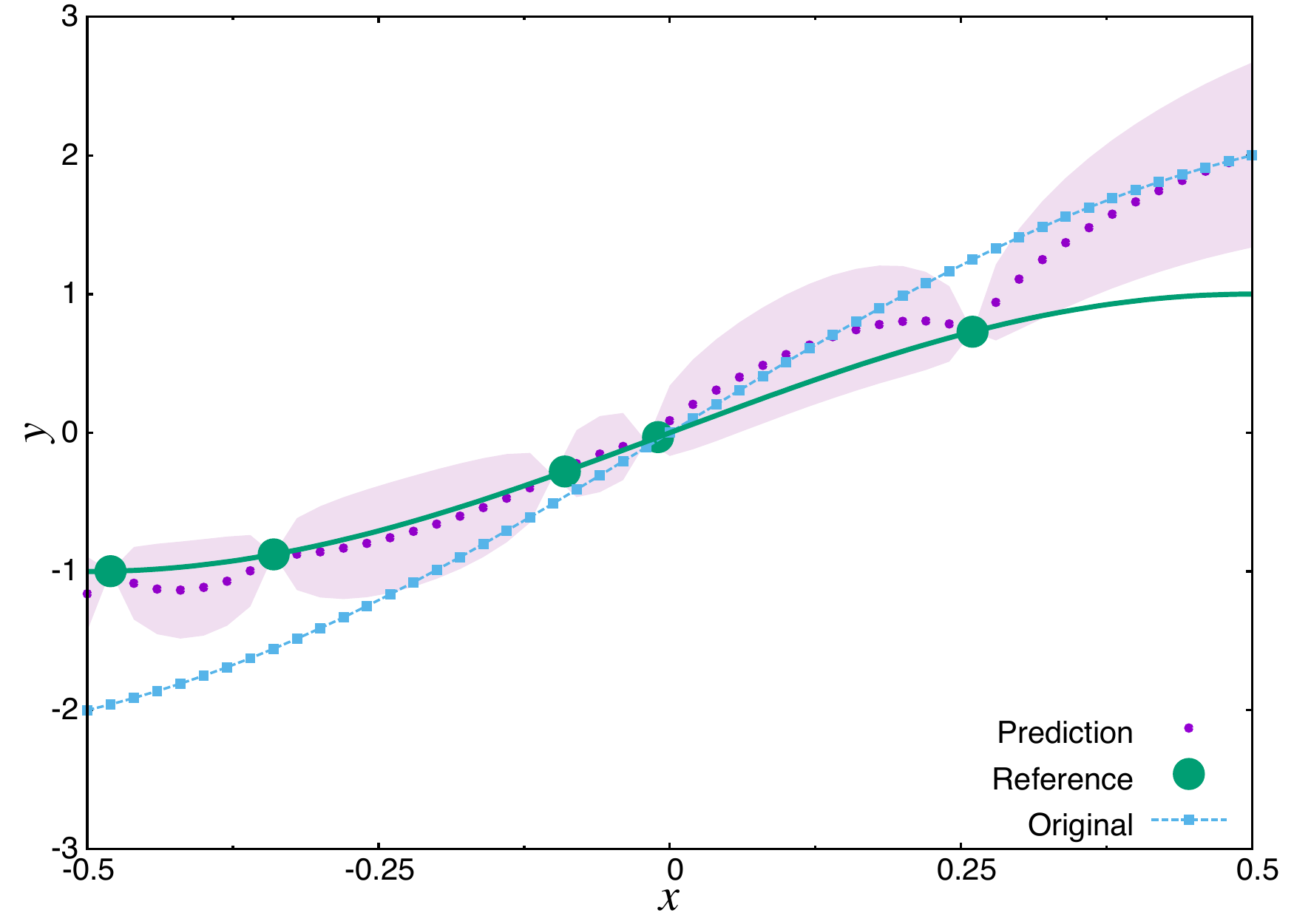} 
 \caption{\label{suriawase_sin2}
 Results of the integration scheme when
 $E[x] = \sin(\pi x)$ (Green line) and
 $E'[x]=\sin(\pi x)+2x$ (cyan points).
 The green circle denote the points of references,
 $E[x_i^\text{R}]$.
 The mean value of
 the improved model, $\tilde{E}[x]$,
 is denoted by the purple points.
 The standard deviation is denoted by the region in light purple.
 }
\end{figure}

The result of the prediction is shown in 
Fig. \ref{suriawase_sin2}.
The original points of $E'[x]$ are shown as
points in cyan.
The purple points show the mean value of 
the model; the light purple region shows the values of 
the standard deviation.
The green curve shows $E[x]$,
the true model, and the green circles are the reference points
used in the prediction.

The mean values, which may be used as 
values of a point estimation,
are improved from the original data
in most of the region.
The error region also covers the 
line of the true model except in the region 
$x>0.26$.

It is noteworthy that 
the bias in the sign of the error mentioned above
makes the uncertainty of the model
large. This would be improved if we assume an asymmetric 
distribution for $\varepsilon_i$ in Eq.~\eqref{model_atomic}.
However,
the resultant form of the integrated model becomes more
complicated.

\begin{figure*}[htbp]
\includegraphics[width=7cm,bb = 0 0 1024 768]{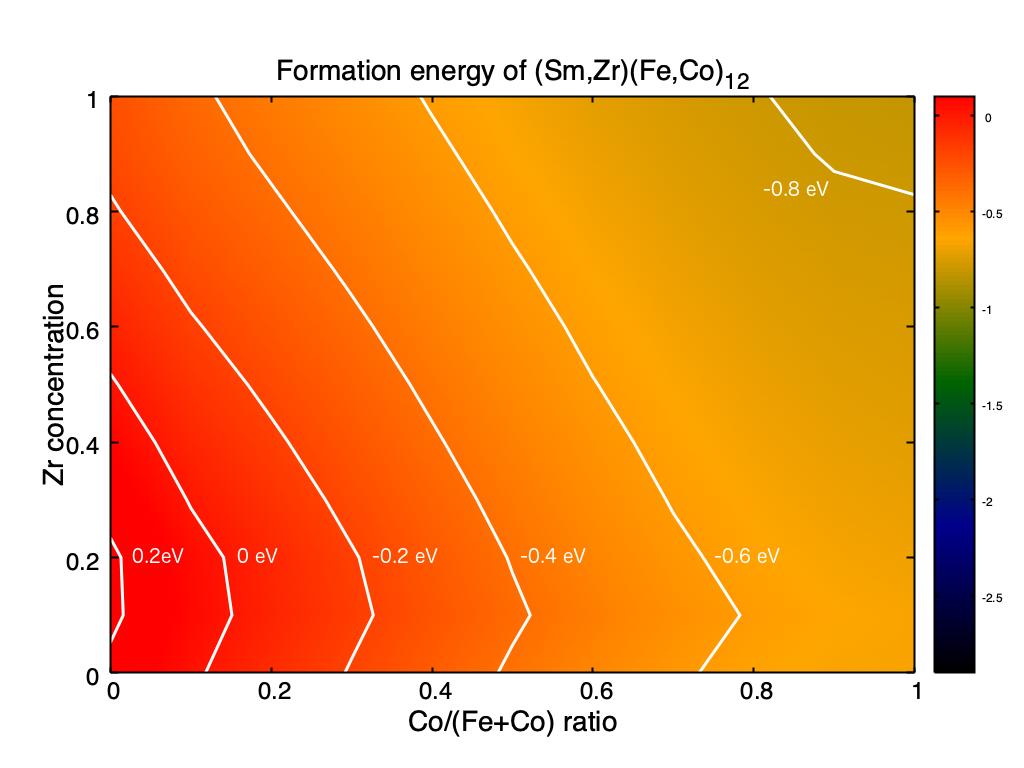} 
\includegraphics[width=7cm,bb = 0 0 1024 768]{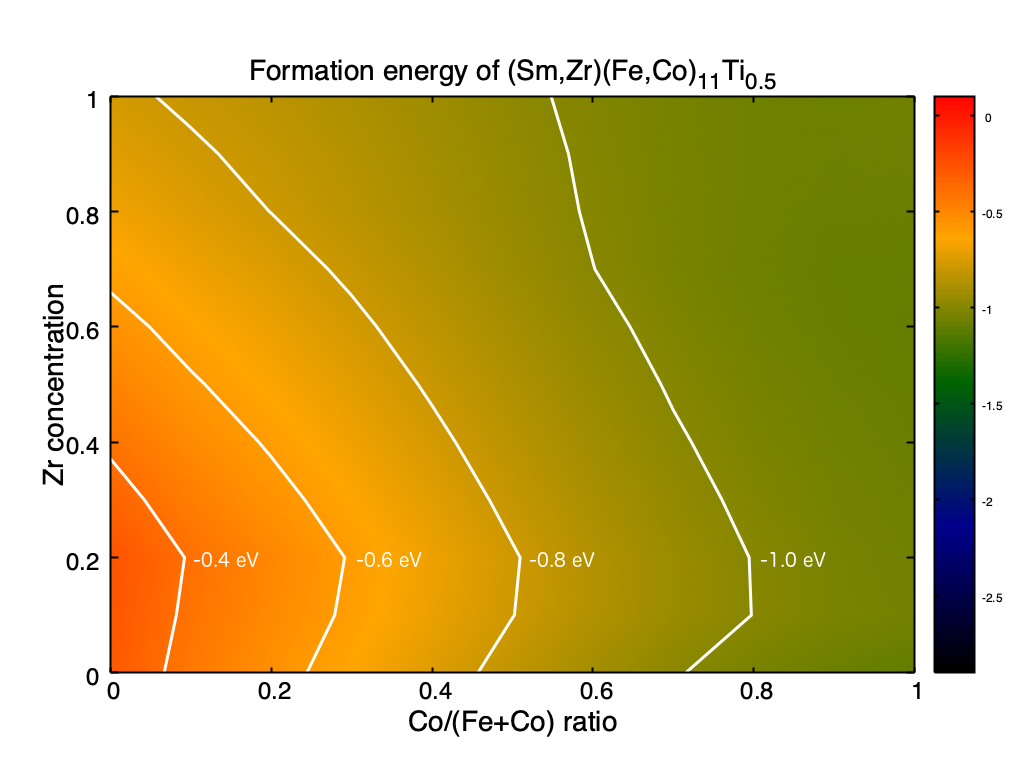} 
\includegraphics[width=7cm,bb = 0 0 1024 768]{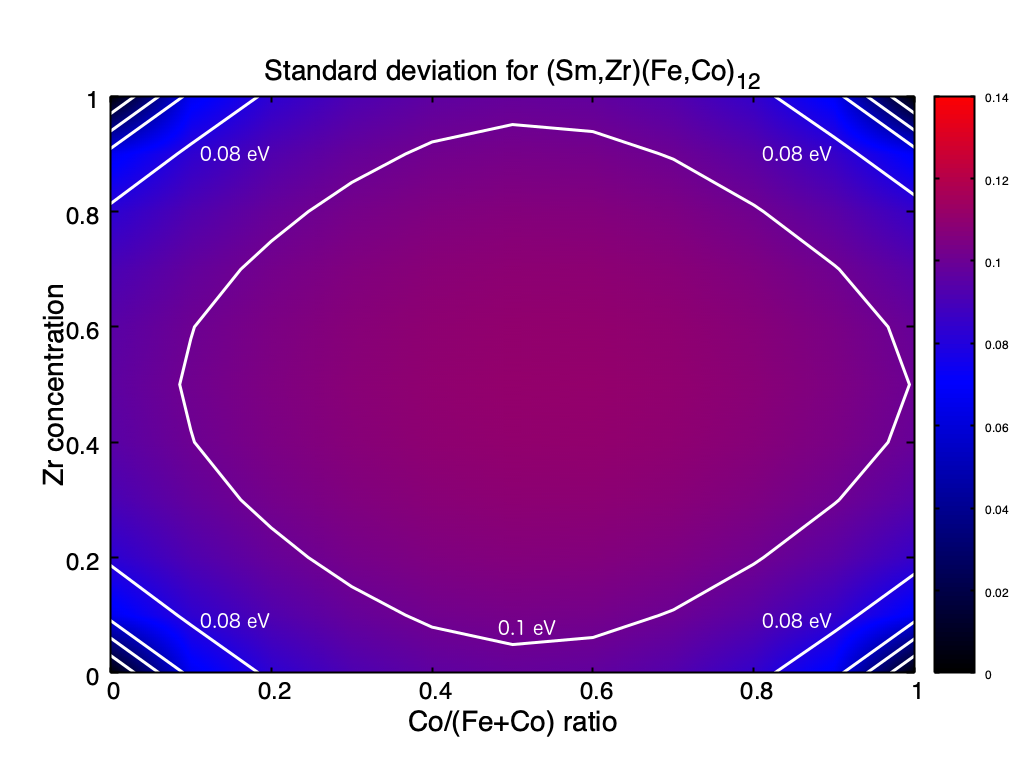} 
\includegraphics[width=7cm,bb = 0 0 1024 768]{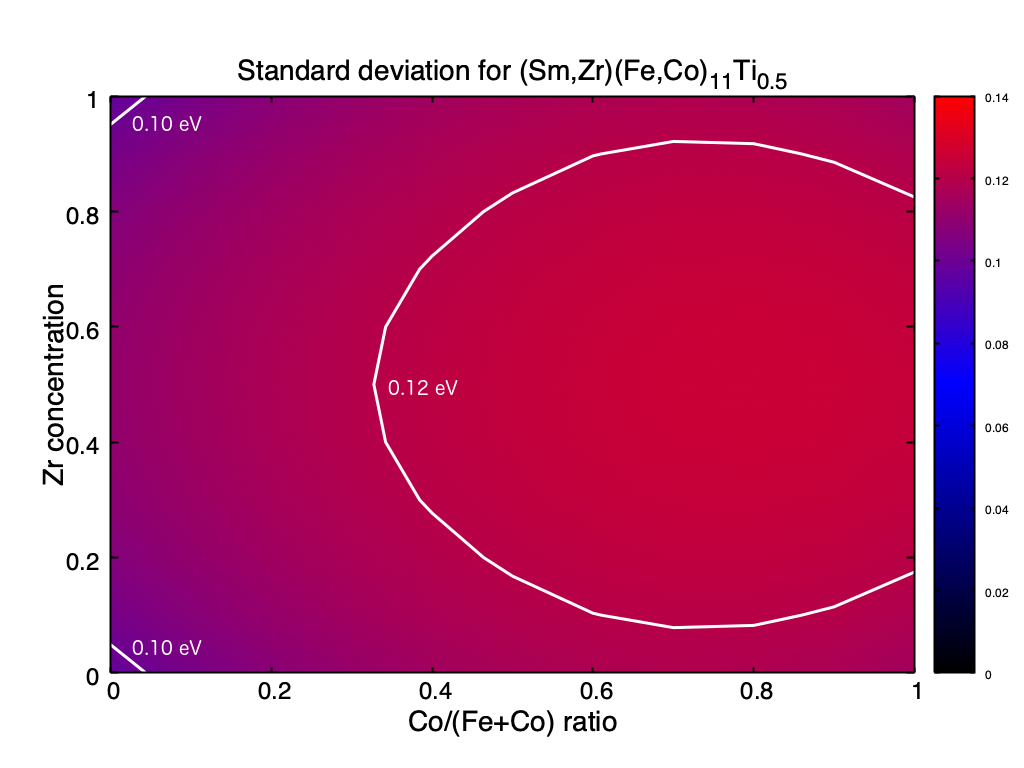} 
 \caption{\label{Interpolated_FE}
 The mean values (top) and the standard deviations (bottom)
 of the integrated model for Sm(Fe,Co)$_{12}$ (left) and 
 Sm(Fe,Co)$_{11.5}$Ti$_{0.5}$ (right).
 The intervals of the contours are 0.2 eV for the mean values
 and 0.02 eV for the standard deviations.
 }
\end{figure*}

Next, let us see
the results of the formation energy
obtained by this method.
Figure~\ref{Interpolated_FE}
shows the formation energy 
of (Sm$_{1-\alpha}$Zr$_\alpha$)
(Fe$_{1-\beta}$Co$_{\beta}$)$_{12-\gamma}$Ti$_{\gamma}$
from the unary systems
for $\gamma = 0$ (left) and $\gamma = 0.5$ (right).
The mean values are shown in the top two figures, and
the standard deviations are shown in the bottom two figures.
In each figures, the horizontal axes show values of
the Co/(Fe+Co) ratio ($\beta$),
and the vertical axes show values of
 the Zr concentration ($\alpha$).

The standard deviation of the model is zero at the corners 
of the figure in $\gamma = 0$ (left bottom)
because they correspond to the reference points,
namely SmFe$_{12}$, ZrFe$_{12}$,
SmCo$_{12}$ and ZrCo$_{12}$.
Therefore, the mean values at the corners of the left top figure
are identical with the reference data (obtained by PAW-GGA).
Because the other reference points are SmFe$_{11}$Ti
and ZrFe$_{11}$Ti (both at $\beta = 0$, $\gamma = 1$),
the contours are placed slightly to the right
in $\gamma = 0$. This is more obvious in the 
plot for $\gamma = 0.5$ (the right bottom panel).

Although we used the mean values ($\tilde{\mu}$) 
in the Bayesian optimization in Section \ref{Sec_BO_results},
it is important to check how the model 
has uncertainty in its prediction, which is 
typically represented by 
the standard deviation $\tilde{S}$:
the model needs more reference points
when $\tilde{S}$ is too large compared with
the value of $\tilde{\mu}$.
One can also make use of
the upper (lower) confidence bound
 $\tilde{\mu}+k\tilde{S}$
($\tilde{\mu}-k\tilde{S}$)
---where $k$ is a positive adjustable parameter---
instead of $\tilde{\mu}$ to 
take account of the uncertainty 
in a pessimistic (optimistic) manner.

\subsection{Bayesian Optimization}
\label{Sec_BO_results}
\begin{figure}[htbp]
\includegraphics[width=8.5cm,bb = 0 0 324 216]{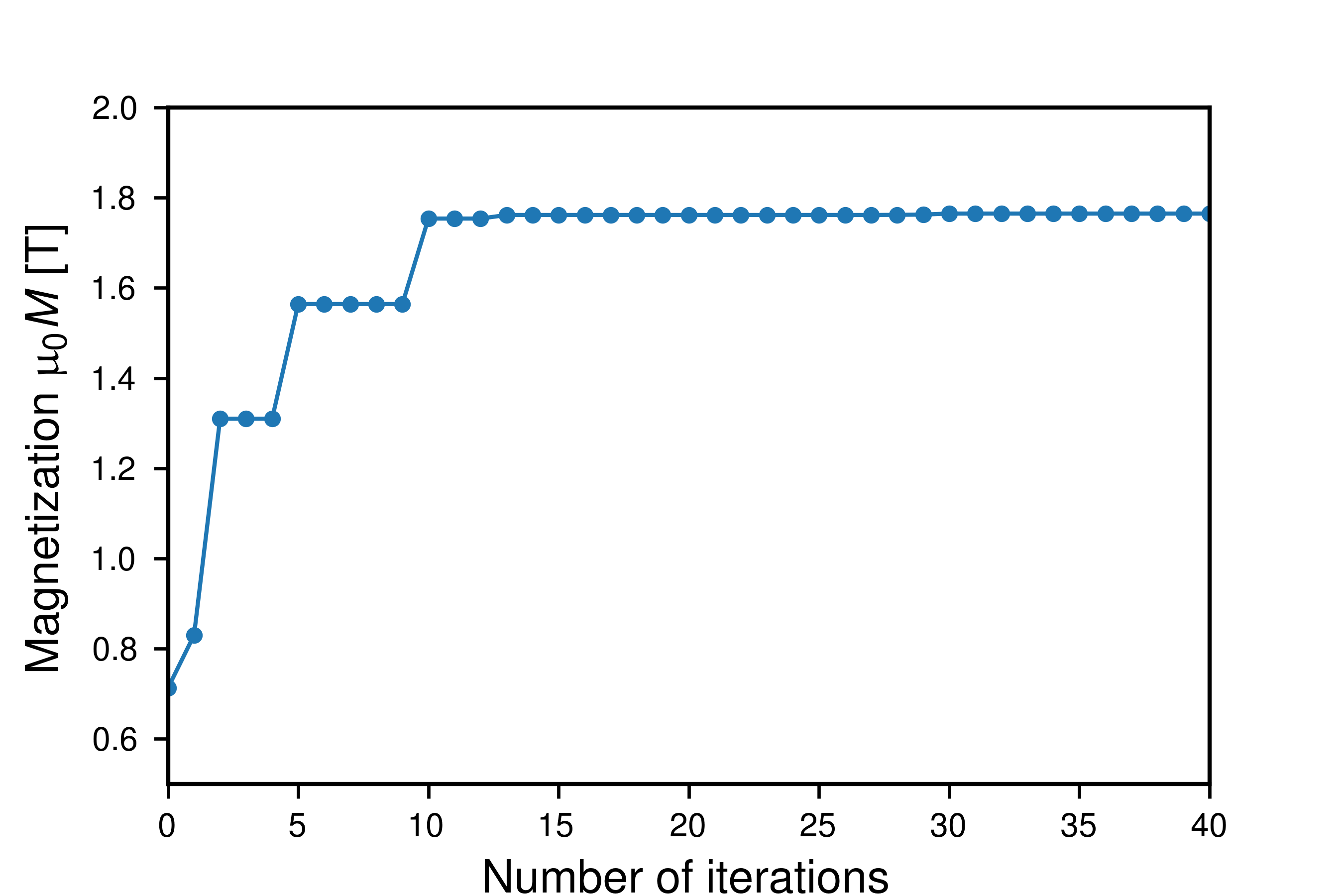}
\caption{
The best magnetization found in the loop of
a session as a function of 
the number of iterations.
The first result in a session corresponds to the point
at the 0th iteration.
\label{fig_one_session}}
\end{figure}

In this subsection, we show the performance of the Bayesian optimization
and discuss the results.
Figure \ref{fig_one_session}
shows one of the optimization processes
with respect to magnetization
using the descriptor \#8.
The highest magnetization found in the first $i$ iterations, 
$\max_{j \leq i} \, \mu_0 M(j) $, is plotted 
against the number of iterations, $i$.
In this run, the highest $\mu_0 M$ in all the 3630 systems was 
found at 30th iteration where we define the zeroth iteration
as that with the first sample.
This process depends on a random sequence which is used
in the sampling from the candidate lists.
In order to take statistical profiles, 
we repeat the optimization scheme (which we call a session)
1,000 times.

\begin{figure*}[htbp]
\includegraphics[width=8.5cm,bb = 0 0 324 216]{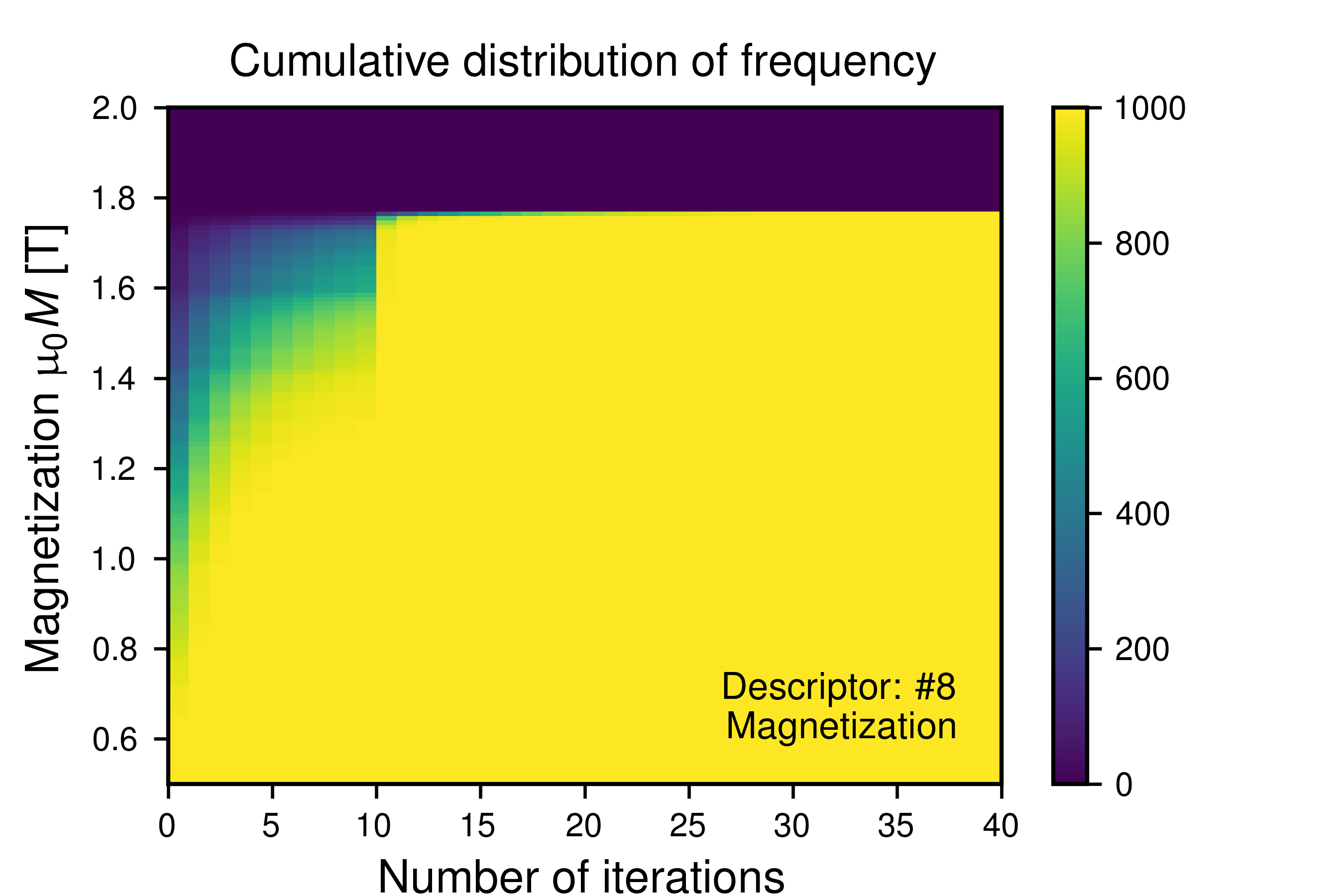} 
\includegraphics[width=8.5cm,bb = 0 0 324 216]{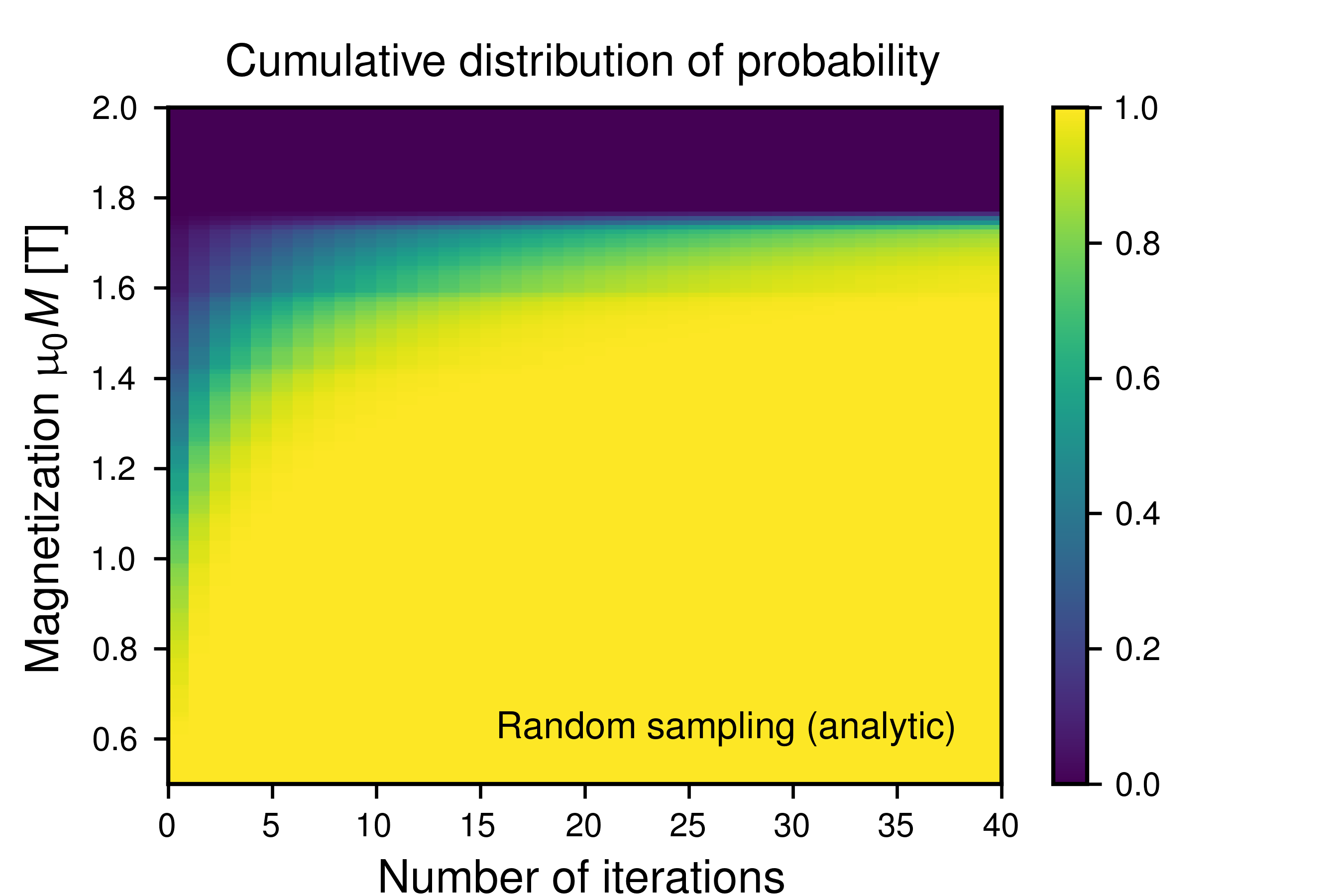} 
 \caption{
 (Left) The cumulative distribution of frequency $\mathcal{D}_i(s)$
 in the optimization of magnetization
 using the Bayesian optimization with the descriptor \#8.
 (Right) The cumulative probability of probability $\mathcal{P}_i(s)$
 that is analytically calculated for the optimization of magnetization
 with the random sampling.
 \label{Cumulative_plots}
 }
\end{figure*}

To analyze the efficiency as a function of the number of iterations,
we consider 
a cumulative distribution $\mathcal{D}_i(s)$
that is defined as
the number of the sessions in which
a system
with a higher score than $s$
is found 
in the first $i$ iterations.
We show a plot of $\mathcal{D}_i(s)$ in Bayesian optimization
of magnetization
using the descriptor \#8 as the left figure 
in Fig. \ref{Cumulative_plots},
where the horizontal axis shows the number of iterations, $i$,
and the vertical axis shows the score variable, $s$.
We also show a plot of the cumulative probability $\mathcal{P}_i(s)$
in the random sampling
that is analytically obtained at the right-hand side.
Because we took an enough number of sessions, there is negligible
difference between the two figures for the first 10 iterations.
The efficiency in the left figure suddenly becomes improved when 
the Bayesian optimization is switched on.

\begin{figure*}[htbp]
\includegraphics[width=12cm,bb = 0 0 504 360]{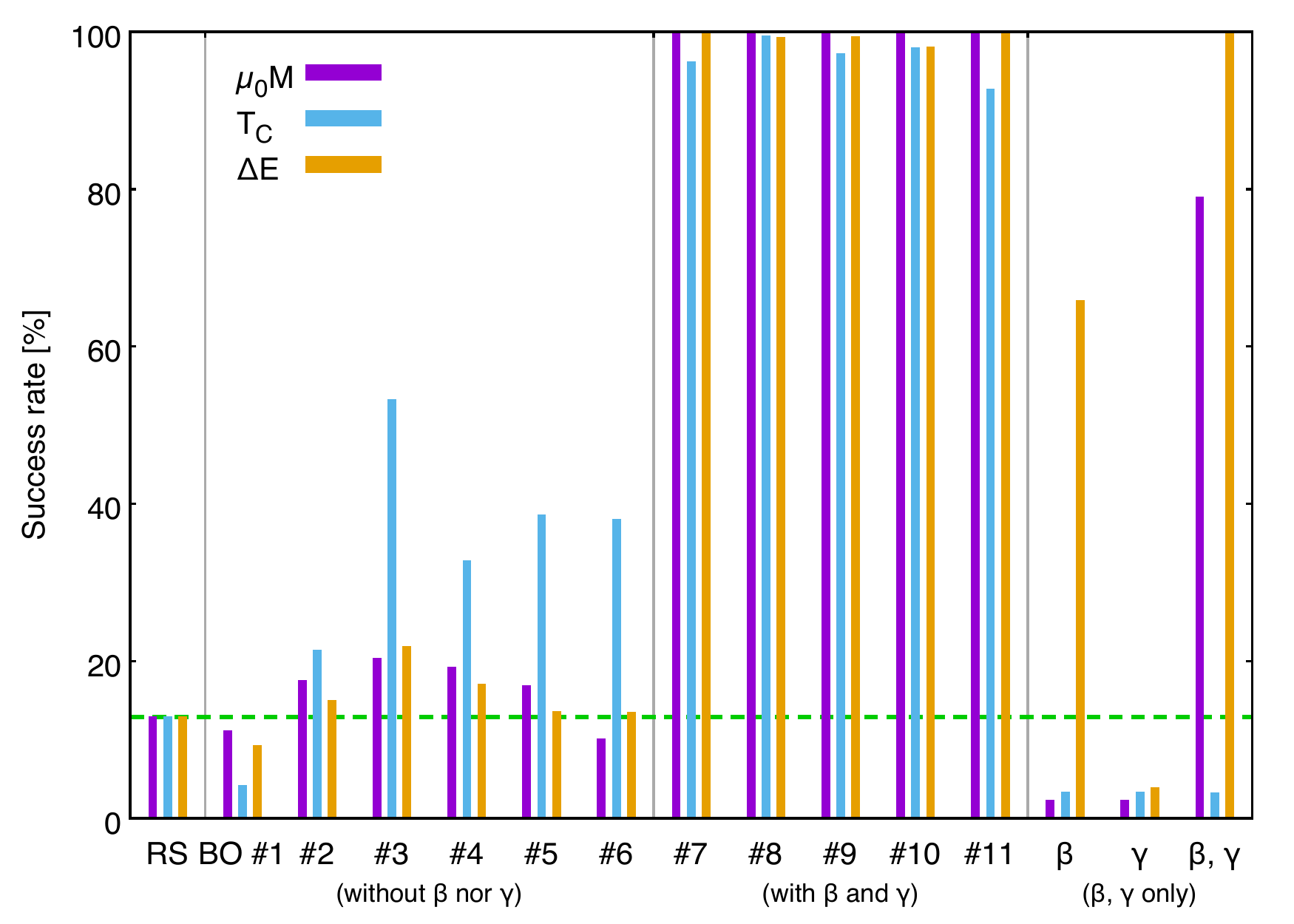} 
\caption{Success rate of finding the systems with the top 10 values of 
magnetization ($\mu_0 M$), Curie temperature ($T_\text{C}$) and formation 
energy from the unary systems ($\Delta E$) among the 3630 candidates
within 50 steps.
RS stands for the random search; BO stands for the Bayesian optimization.
The descriptors used in the search are shown along the horizontal axis:
the numbers denote the descriptors listed in Table \ref{tab_descriptors};
the label $\beta$, $\gamma$ and $\beta, \gamma$ denote the results with using 
$\beta$, $\gamma$ and $\beta, \gamma$ as descriptors, respectively.
The horizontal dashed line (green) is an eye-guide to show the height of the bars for the random sampling.
\label{fig_top10_50}}
\end{figure*}
Figure \ref{fig_top10_50} shows the success rate of finding the systems 
with the top 10 values of the target properties---magnetization ($\mu_0 M$),
Curie temperature ($T_\text{C}$) and formation 
energy from the unary systems ($\Delta E$)---within 50 steps.
The results with Bayesian optimization (BO) are compared with the search by the random sampling (RS).
The numbers with \# in the figure denote the 
descriptors listed in Table \ref{tab_descriptors}.
We find that the efficiency significantly depends on the choice of the descriptor. 
It is obvious from the figure that 
the descriptors \#7--11 are very efficient in Bayesian optimization and much superior to 
those with \#1--6 and the random sampling.
This clearly shows that $\beta$ and $\gamma$ are important factors
because the descriptors \#7--11 differs from \#2--6 only by 
$\beta$ and $\gamma$ used instead of $N_\mathrm{T}$.

This example demonstrates how we can incorporate domain knowledge into the machine learning.
It is known that 
the magnetization of ferromagnetic random alloys of Fe and another transition metal
is usually a well-behaved function of the number of the valence electrons. 
It is called Slater-Pauling curve.
This curve can be reproduced well by first-principles calculation
with CPA\cite{Dederichs91}, and so is the Slater-Pauling-like 
curve for the Curie temperature.\cite{Takahashi07}
These effects 
have also been observed
in ThMn$_{12}$-type compounds
experimentally,\cite{Buschow91, Hirayama17}
and explained theoretically.\cite{Miyake14,Harashima15e,Harashima16,Fukazawa18,Harashima18}
On the basis of these previous researches, we were able to expect that
including $\beta$ [the Co/(Fe+Co) ratio]
and $\gamma$ (the Ti content) into the descriptor
would improve the efficiency of the search in advance.

However, we also find that $\beta$ and $\gamma$ alone do not work
as an efficient descriptor.
The results of the Bayesian optimization with using $\beta$, $\gamma$, and 
the pair of them as descriptors are also shown in 
Fig.~\ref{fig_top10_50}.
Those success rates are significantly lower than the rates with 
the descriptors \#7--11.
This is because there are 66[= 3 (for R) $\times$ 2 (for Z) $\times$ 11
(for $\alpha$)] candidates that has 
common values of $\beta$ and $\gamma$ on the list, and
50 steps are not enough to obtain an adequate Bayesian model
and draw one of the top 10 systems by chance.
It is noteworthy that the efficiency of
the Bayesian optimization is largely improved
by adding only $N_\mathrm{2a}$
as in the descriptor \#7.

\begin{figure}[htbp]
\includegraphics[width=7cm,bb = 0 0 324 216]{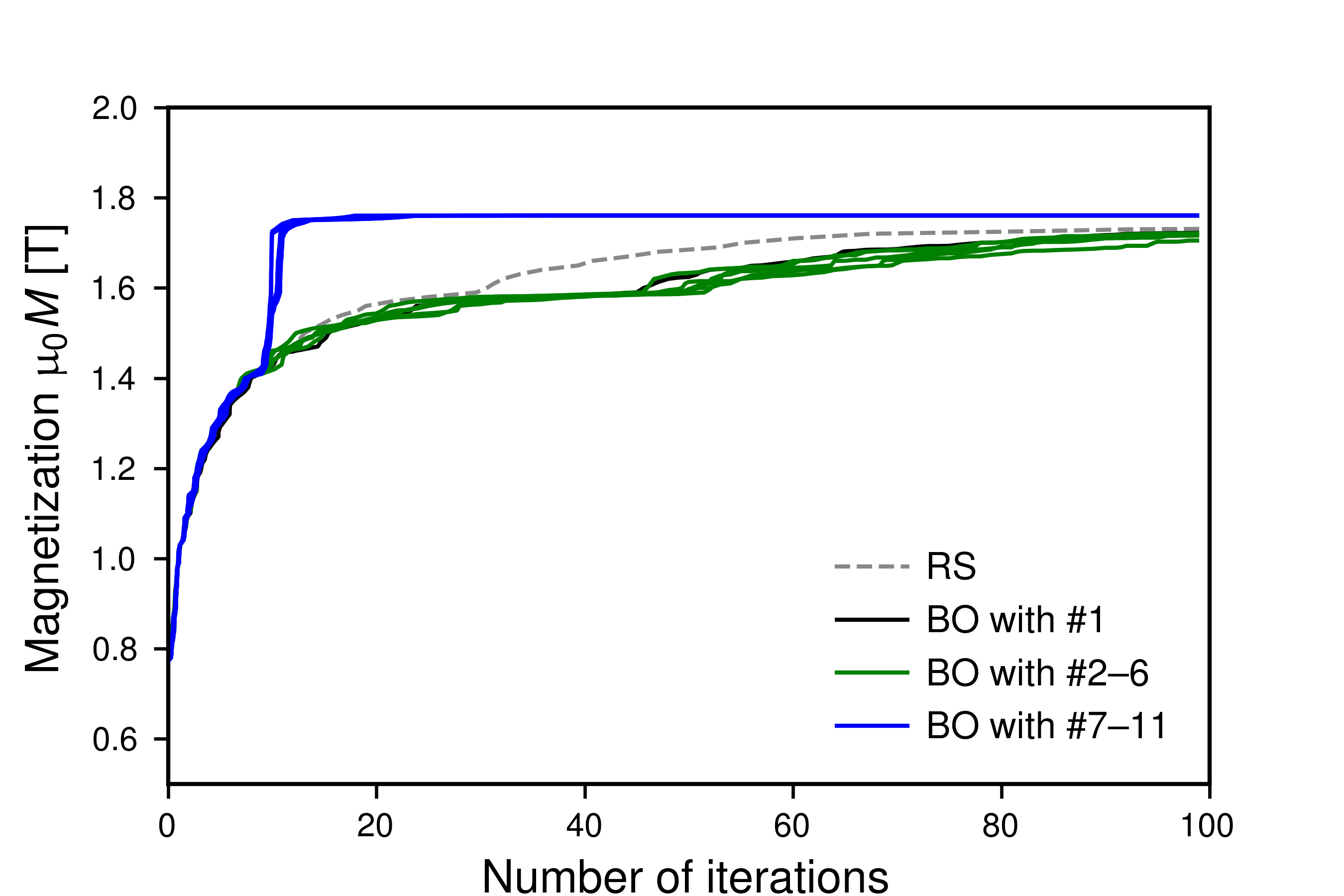} 
\includegraphics[width=7cm,bb = 0 0 324 216]{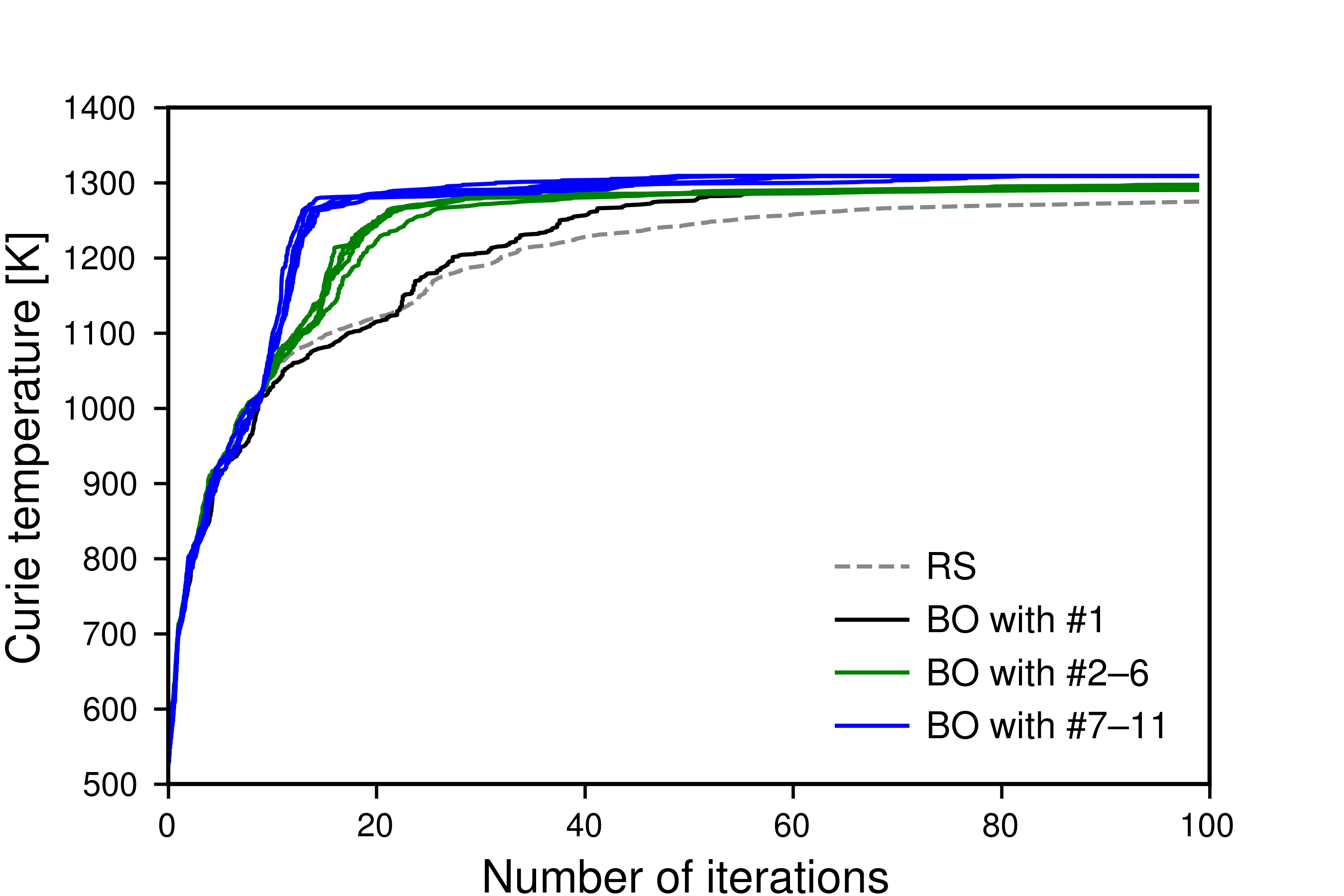} 
\includegraphics[width=7cm,bb = 0 0 324 216]{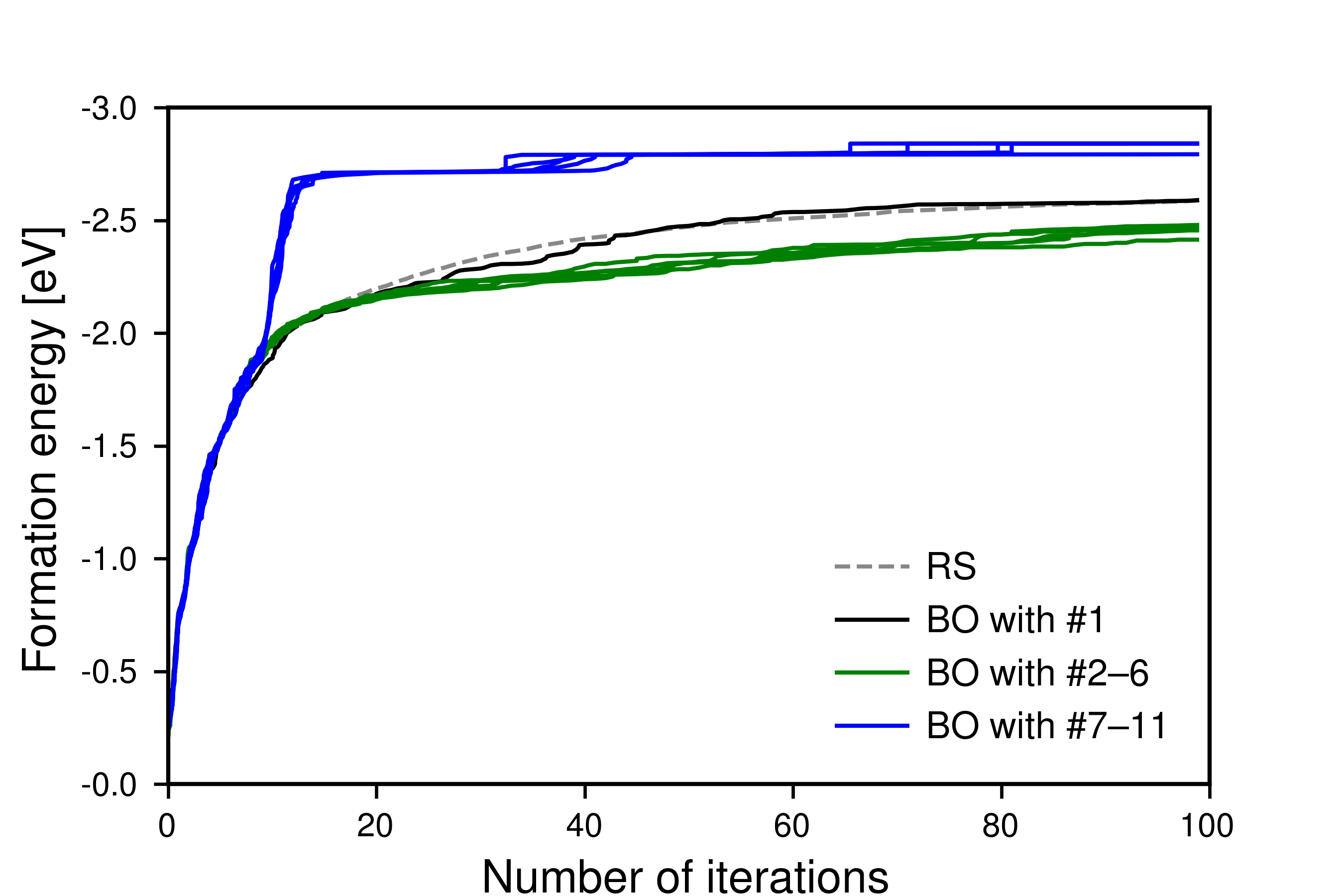} 
 \caption{\label{Fig_score_vs_steps}
 The 95\% contours of the cumulative
 distribution $D_i(s)$
 in the random sampling (gray dashed)
 and in the Bayesian optimizations with
 the descriptor \#1 (black),
 the descriptors \#2--6 (green)
 and  the descriptors \#7--11 (blue).
 The horizontal axis shows $i$; the vertical axis shows $s$.
 The top, middle and bottom figures show results in the optimization
 of the magnetization, 
 the Curie temperature within the mean-field approximation,
 and 
 the formation energy from the unary systems, respectively.
 }
\end{figure}

Figure \ref{Fig_score_vs_steps} shows the 95\% (950 sessions) contour
of $D_i(s)$ obtained with the descriptors \#1--11.
The best 95\% of the data points are laid above those curves.
The panels show results with the random search and the
Bayesian optimization with the descriptors \#1--11.
As shown in these graphs, the efficiency of the search depends
much on the target property to optimize.
The descriptors \#7--11, which are with $\beta$ and $\gamma$,
has a satisfying efficiency, with which even 20 iterations are
enough to obtain a nearly best score regardless of the
target property.
The situation is quite different
with the descriptors \#1--6.
In the optimization of the Curie temperature,
these descriptors works efficiently.
However, the optimizations of the magnetization and 
the formation energy progress even more slowly than
the random sampling.

\begin{figure}[htbp]
\includegraphics[width=7cm,bb = 0 0 324 216]{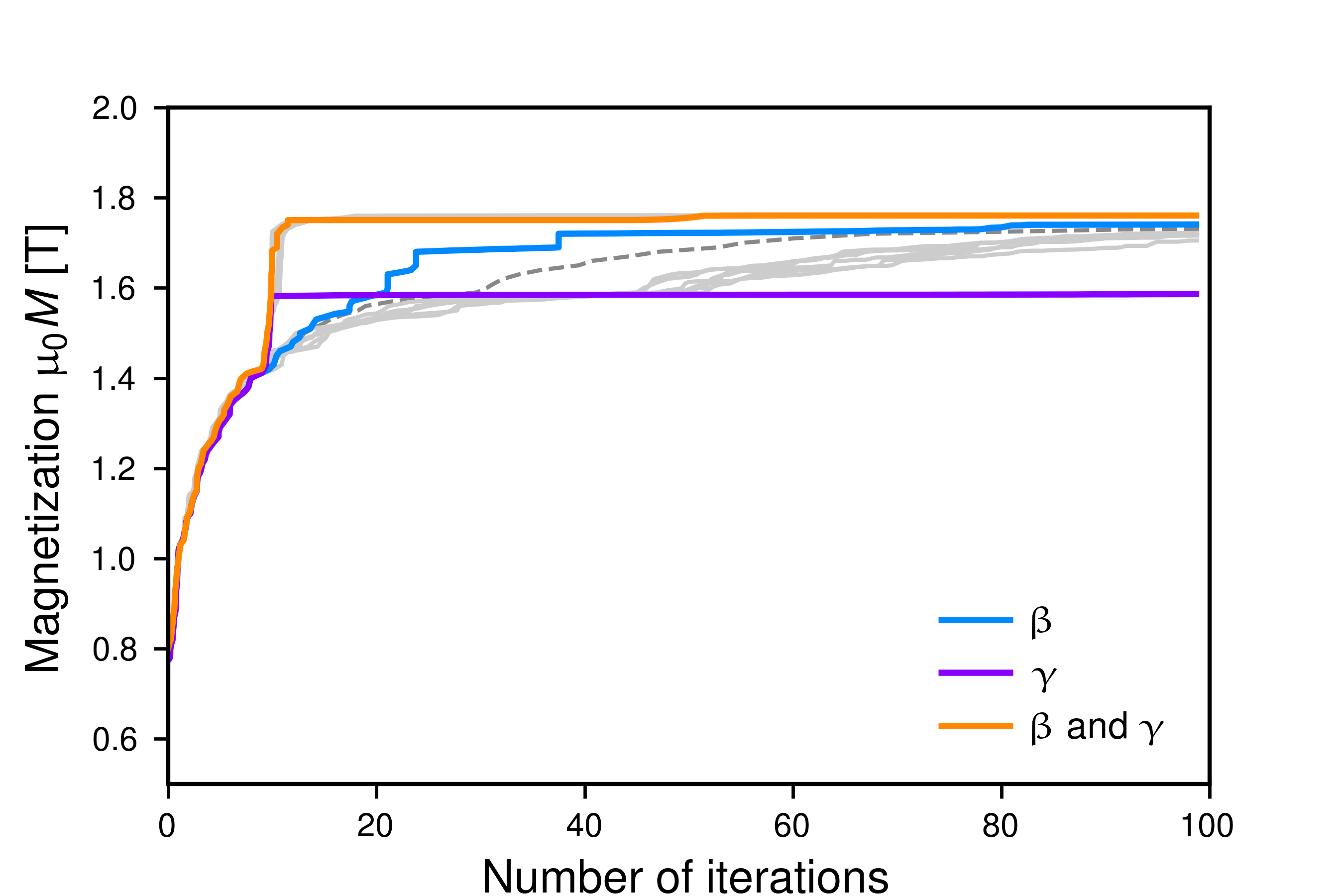} 
\includegraphics[width=7cm,bb = 0 0 324 216]{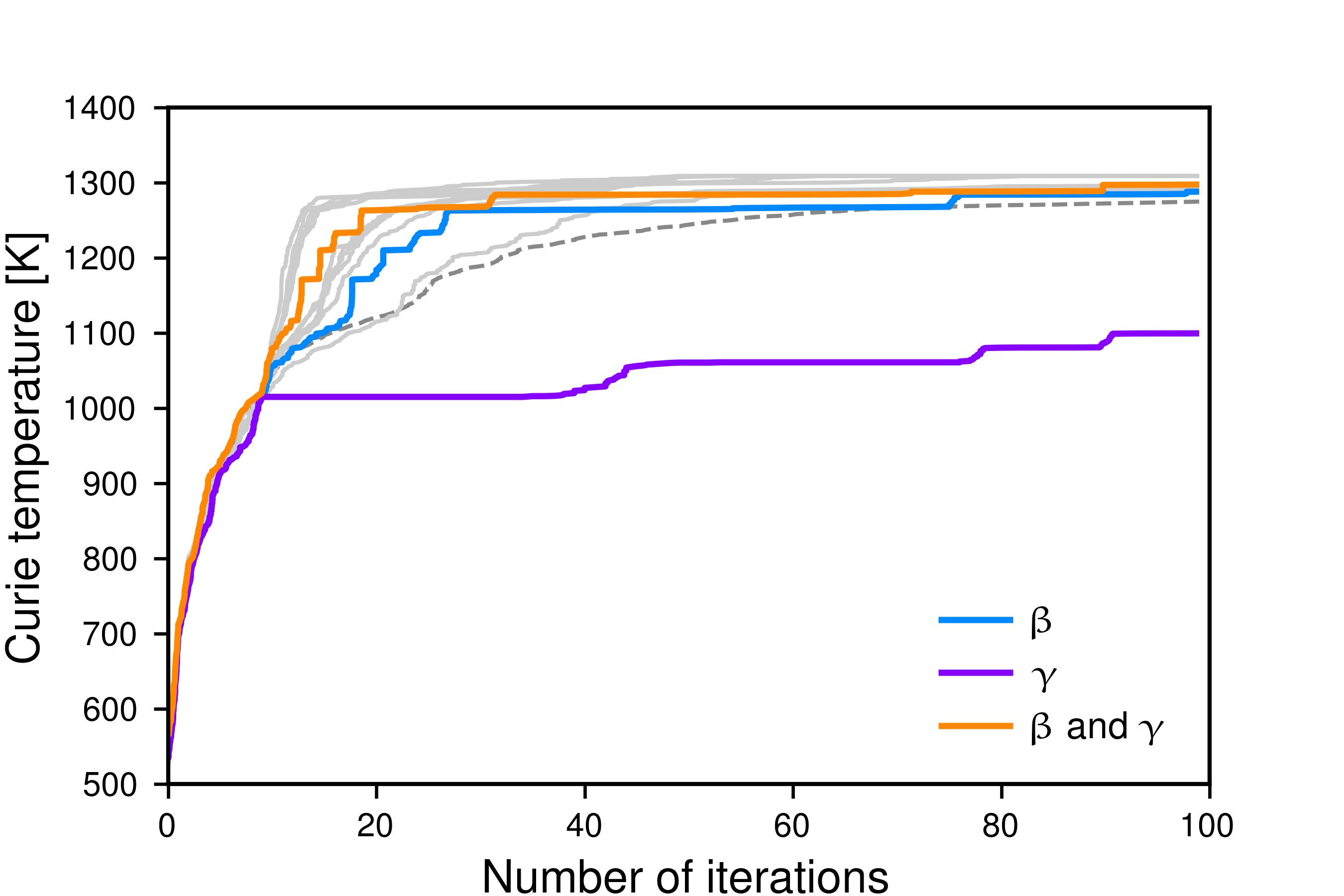} 
\includegraphics[width=7cm,bb = 0 0 324 216]{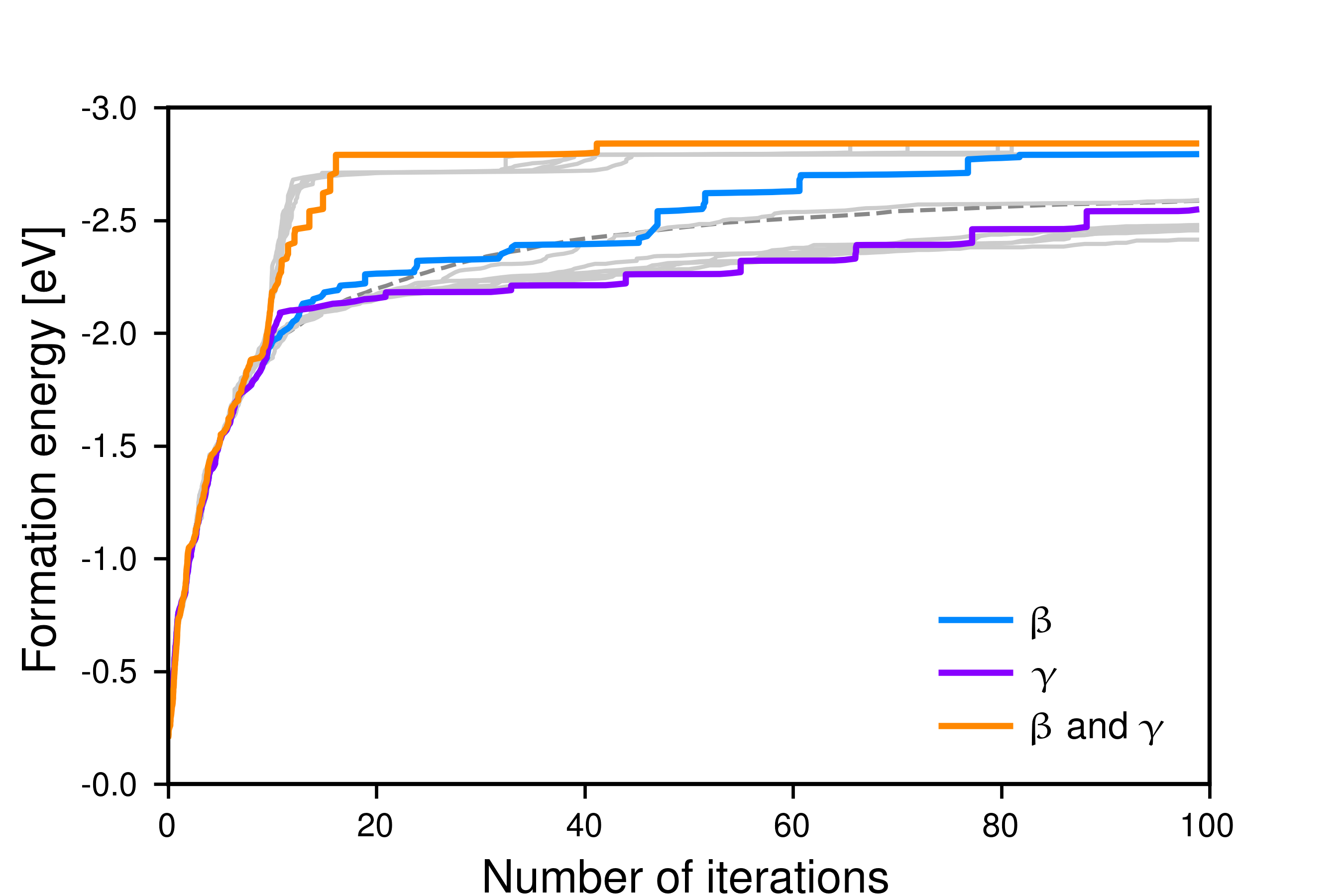} 
 \caption{\label{Fig_score_vs_steps_2}
 The 95\% contours of the cumulative
 distribution $D_i(s)$
 in the Bayesian optimizations with 
 the descriptor $\beta$ (cyan),
 the descriptor $\gamma$ (purple)
 and the descriptor $(\beta, \gamma)$ (orange).
 Those in the random sampling and in the Bayesian 
 optimization with the other descriptors are also shown
 as gray lines.
 The horizontal axis shows $i$; the vertical axis shows $s$.
 The top, middle and bottom figures show results in the optimization
 of the magnetization, 
 the Curie temperature within the mean-field approximation,
 and 
 the formation energy from the unary systems, respectively.
 }
\end{figure}

This difference between the Curie temperature
and the other targets
is also seen
in Fig.~\ref{Fig_score_vs_steps_2}, where
the pair of $\beta$ and $\gamma$ is used
as descriptors.
The pair descriptor works as efficient as 
the descriptors \#7--11 in the optimization 
of the magnetization and the formation energy.
However, discernible difference in efficiency 
exists in the optimization of the Curie temperature.
This suggests importance of information about elements
at the 2a site (R and Z), which is consistent with
Dam et al's observation that the concentration of 
rare-earth elements is important in explaining
the Curie temperature of binary alloys
composed of a rare-earth element and 
a 3d transition-metal.\cite{Dam18}

Dependency of the search efficiency
on the the dimension of the descriptor
is also noteworthy.
When the dimension of a descriptor is large,
the descriptor can accommodate a large search space on the one hand.
However, modeling tends to be difficult with a higher dimensional 
space, which is referred
as ``curse of dimensionality'', on the other hand.
Note that 
we have descriptors with 4 different dimensions
in the groups of 
the descriptors \#1--6 and \#7--11.
Figure~\ref{Fig_score_vs_steps} shows that the dimension 
has only a minor effect on the efficiency.
This dependence is magnified 
by the more stringent criterion of the top-10 benchmark
as shown in Fig.~\ref{fig_top10_50},
especially in
the results with the descriptor \#1--6.
However, the difference among the descriptor \#7--11 are still subtle.
Therefore, we expect that
it would be safe to include six or a little
more variables into the descriptor
when we optimize for another target property.

\section{Conclusion}
\label{Conclusion}
In this paper, we presented a machine-learning scheme
for searching high-performance magnetic compounds. 
Our scheme is based on Bayesian optimization,
and has a much higher efficiency
than random sampling.
We demonstrated its efficiency by taking the example of
optimization of
magnetization, Curie temperature and 
formation energy
for the search space of magnet compounds having
the ThMn$_{12}$ structure.
One of the typical results is the success rate of finding
the top 10 systems with the highest properties
when 50 systems are sampled from a candidate list of 3630 systems
(Fig.~\ref{fig_top10_50}).
The success rate is more than 90\% with our scheme
when the descriptor is appropriately chosen
while it is approximately 10\% in the random sampling.

The efficiency is maximized when we include 
the Co content ($\beta$), the Ti content ($\gamma$),
and the information of the R and Z elements
(e.g. $N_\mathrm{2a}$)
into the descriptor.
This improvement is what we could expect from the previous 
studies of magnet compounds.
We stressed that it is important to incorporate domain knowledge
into the choice of a descriptor.
We also discussed how many variables
a descriptor can accommodate
without deteriorating the search efficiency.
Although an excessive addition of variables to the descriptor
can lose the efficiency of the search,
there was not a significant loss
when we doubled the dimension of the optimal descriptor.

We also proposed an integration scheme of two datasets to improve accuracy of
an inexpensive large-sized dataset
with using an accurate and small-sized dataset
(reference dataset).
The algorithm (Alg.~\ref{Algorithm_1} and \ref{Algorithm_2}) is easy to
implement and fast.
Prediction with a confidence bound (or the standard deviation
$\tilde{S}$) is another feature of the scheme.
We have also shown
how it worked in the calculation of the formation energy
(Fig.~\ref{Interpolated_FE}).

\section*{Acknowledgment}
We acknowledge the support from the Elements Strategy Initiative Project under the auspices of MEXT. This work was also supported by MEXT as a social and scientific priority issue (Creation of new functional Devices and high-performance Materials to Support next-generation Industries; CDMSI) to be tackled by using the post-K computer, 
by the ``Materials research by Information Integration'' Initiative
(MI$^2$I) project of the Support Program for Starting Up Innovation Hub from the Japan Science and Technology Agency (JST). 
The computation was partly conducted using the facilities of the Supercomputer Center, the Institute for Solid State Physics, the University of Tokyo, and the supercomputer of ACCMS, Kyoto University. 
This research also used computational resources of the K computer provided by the RIKEN Advanced Institute for Computational Science through the HPCI System Research project (Project ID:hp170100).

\appendix
\section{Lattice parameters}
\label{lattparams}
Table \ref{RCo12_lattice}
lists the lattice parameters for 
RCo$_{12}$
we used in the calculations.
We assumed the ThMn$_{12}$ structure [space group: I4/mmm (\#139)]
for the system.
The definitions of $p_\text{8i}$ and $p_\text{8j}$ are summarized in 
Table \ref{table1} with representable atomic positions of the atoms.
\begin{table}[htbp]
  \caption{
 Optimized lattice parameters for RCo$_{12}$ (R = Y, Nd, Sm).
 See Table \ref{table1} for definitions of $p_{\rm 8i}$ and $p_{\rm 8j}$.
  \label{RCo12_lattice}}
 \begin{tabular}{ccccc}
  \hline
  \hline
   R & $a$ [\AA] & $c$ [\AA] &$p_{\rm 8i}$& $p_{\rm 8j}$\\
  \hline
                          Y &  8.282 &   4.659 & 0.3585 & 0.2738 \\
                         Nd &  8.336 &   4.677 & 0.3590 & 0.2695 \\
                         Sm &  8.309 &   4.669 & 0.3587 & 0.2715 \\
  \hline
  \hline
 \end{tabular}
 \end{table}
\begin{table}[htbp]
 \caption{Representable atomic positions in the ThMn$_{12}$ structure. The variables, $x$, $y$, and $z$, denote the point ($ax$, $ay$, $cz$) in Cartesian coordinates.
 \label{table1}
 }
 \begin{tabular}{ccrrr}
  \hline
  \hline
  Element & Site & $x$ & $y$& $z$\\
  \hline
  Th & 2a &            0 &    0 &    0 \\
  Mn & 8f &         0.25 & 0.25 & 0.25 \\
  Mn & 8i & $p_{\rm 8i}$ &    0 &    0 \\
  Mn & 8j & $p_{\rm 8j}$ &  0.5 &    0 \\
  \hline
  \hline
 \end{tabular}
 \end{table}

In our KKR-CPA calculations
for RFe$_{11}$Ti (R=Y, Nd, Sm) 
and ZFe$_{11}$Ti (Z=Zr, Dy),
we assumed that they also have the ThMn$_{12}$ structure.
The lattice parameters are reduced from
the structure obtained in Ref. \onlinecite{Harashima14b}
and the structure given in Table \ref{Table_ZFe11Ti}.
The value of $\sqrt{ab}$ is used as the $a$ parameter for the 
reduced cell to keep the cell volume.
The inner parameters are determined so that 
the they minimize the deviation in the space of
the coefficients $(x,y,z)$ of the unit vectors,
which corresponds to the point ($ax$,$by$,$cz$) in
the Cartesian coordinates.
The reduced values are listed in 
Table~\ref{ZFe11Ti_lattice}.

\begin{table*}[htbp]
  \caption{Inner coordinates $(x, y, z)$
  of Fe and Ti in DyFe$_{11}$Ti and ZrFe$_{11}$Ti
  where Dy and Zr are placed at (0, 0, 0). The position of Ti is denoted by 
  a parenthesis.
  These values corresponds 
  to the point ($ax$, $by$, $cz$) in Cartesian coordinates where
  $a=8.455$, $b=8.262$ and $c=4.715$ in ZrFe$_{11}$Ti, and
  $a=8.518$, $b=8.410$ and $c=4.727$ in DyFe$_{11}$Ti.
 \label{Table_ZFe11Ti}} 
  \vspace{5pt}
  \begin{tabular}{lrrr|rrr|rrr}
    \hline
    \hline
    & \multicolumn{3}{c|}{Fe(8f)} & \multicolumn{3}{|c|}{Fe(8i)} & \multicolumn{3}{|c}{Fe(8j)} \\ 
    & \multicolumn{1}{c}{$x$} & \multicolumn{1}{c}{$y$} & \multicolumn{1}{c|}{$z$}
    & \multicolumn{1}{c}{$x$} & \multicolumn{1}{c}{$y$} & \multicolumn{1}{c|}{$z$}
    & \multicolumn{1}{c}{$x$} & \multicolumn{1}{c}{$y$} & \multicolumn{1}{c} {$z$} \\
    \hline
    DyFe$_{11}$Ti
    &\ $ 0.255$ & $ 0.251$ & $ 0.251$ &  ($  0.375$ & $  0.000$ & $ 0.000$)&  $  0.272$ & $  0.500$ & $ 0.000$  \\
    &  $ 0.255$ & $ 0.749$ & $ 0.749$ &\  $ -0.348$ & $  0.000$ & $ 0.000 $&\ $ -0.266$ & $  0.500$ & $ 0.000$  \\
    &  $ 0.755$ & $ 0.249$ & $ 0.751$ &   $  0.005$ & $  0.355$ & $ 0.000 $&  $  0.506$ & $  0.286$ & $ 0.000$  \\
    &  $ 0.755$ & $ 0.751$ & $ 0.249$ &   $  0.005$ & $ -0.355$ & $ 0.000 $&  $  0.506$ & $ -0.286$ & $ 0.000$  \\
    \hline
    ZrFe$_{11}$Ti
    & $ 0.256$ & $ 0.250$ & $ 0.251$ &   ($  0.381$ & $  0.000$ & $ 0.000$)&  $  0.276$ & $  0.500$ & $ 0.000$  \\
    & $ 0.256$ & $ 0.750$ & $ 0.750$ &    $ -0.342$ & $  0.000$ & $ 0.000 $&  $ -0.264$ & $  0.500$ & $ 0.000$  \\
    & $ 0.756$ & $ 0.250$ & $ 0.751$ &    $  0.007$ & $  0.352$ & $ 0.000 $&  $  0.507$ & $  0.300$ & $ 0.000$  \\
    & $ 0.756$ & $ 0.750$ & $ 0.250$ &    $  0.007$ & $ -0.352$ & $ 0.000 $&  $  0.507$ & $ -0.300$ & $ 0.000$  \\
    \hline
    \hline
  \end{tabular}
\end{table*}
\begin{table}[htbp]
  \caption{
 Reduced values of the lattice parameters for RFe$_{11}$Ti (R = Y, Nd, Sm)
 and ZFe$_{11}$Ti (R = Zr, Dy).
 See Table \ref{table1} for definitions of $p_{\rm 8i}$ and $p_{\rm 8j}$.
  \label{ZFe11Ti_lattice}}
 \begin{tabular}{ccccc}
  \hline
  \hline
   Z & $a$ [\AA] & $c$ [\AA] &$p_{\rm 8i}$& $p_{\rm 8j}$\\
  \hline
                          Y &  8.476 &   4.730 & 0.3606 & 0.2764 \\
                         Nd &  8.560 &   4.701 & 0.3596 & 0.2703 \\
                         Sm &  8.523 &   4.713 & 0.3590 & 0.2728 \\
                         Zr &  8.358 &   4.715 & 0.3565 & 0.2850 \\
                         Dy &  8.464 &   4.727 & 0.3584 & 0.2776 \\
  \hline
  \hline
 \end{tabular}
\end{table}

\section{Integration model with a linear term}
\label{Model_with_Linear_term}
In this section, we incorporate an adjustable 
linear term into
the relation of $E[\vec{x}]$ and $E'[\vec{x}]$,
which appear
in Section \ref{Suriawase}:
we consider 
$E'[\vec{x}]$ as an approximate function of
$E[\vec{x}] + \vec{a}\cdot\vec{x} + b$ in this section.

This changes Eq.~\eqref{model_atomic} to
\begin{equation}
 \tilde{E}_i[\vec{y}]
  -
  E[\vec{x}^\text{\,R}_i]
  =
  E'[\vec{y}]
  -
  E'[\vec{x}^\text{\,R}_i]
  +
  \vec{a}\cdot\left(\vec{y} - \vec{x}^\text{\,R}_i \right)
  +
  \varepsilon_i.
\end{equation}
Therefore, our model does not depend on the variable $b$.

Equation \eqref{Suriawase_mu} is modified as follows:
\begin{equation}
 \tilde{\mu}
  =
  E'[\vec{y}]
  +
  \frac{1}{\Omega}
  \sum_i
  \omega_i
  \left\{
      E[\vec{x}^\text{\,R}_i]
      -
      E'[\vec{x}^\text{\,R}_i]
      +
      \vec{a}\cdot\left(\vec{y} - \vec{x}^\text{\,R}_i \right)
  \right\},
\end{equation}
while Eq.~\eqref{Suriawase_S2} is left unchanged.

The estimation
by the maximum likelihood method with $\tilde{E}_{\text{LOO},i}$
described in Section \ref{Suriawase}
is applicable to the variable $\vec{a}$ and $\sigma^2$.
The equation for $\vec{a}$ is 
\begin{widetext}
\begin{equation}
 \left(
 \sum_i
  \frac{1}
  {\Omega_{\text{LOO},i}}
  \vec{\xi}_i \otimes \vec{\xi}_i
 \right)
 \vec{a}
 =
 \sum_i
 \left(
    E[\vec{x}^\text{\,R}_i]
    -
    E'[\vec{x}^\text{\,R}_i]
    -
    \frac{1}
    {\Omega_{\text{LOO},i}}
    \sum_{k\neq i}
    \omega_k
    \left\{
        E[\vec{x}^\text{\,R}_k]
        -
        E'[\vec{x}^\text{\,R}_k]
    \right\}
 \right)
\end{equation}
\end{widetext}
where $\vec{\xi}_i$ is defined as
\begin{equation}
 \vec{\xi}_i
  \equiv
  \sum_{k\neq i}
  \omega_k
  \{
      \vec{x}^\text{\,R}_i - \vec{x}^\text{\,R}_k
  \}
\end{equation}
and the symbol $\otimes$ denotes the dyadic product, with which 
$(\alpha_1, \alpha_2, \alpha_3)\otimes(\beta_1, \beta_2, \beta_3)$
is defined as the matrix whose $(i,j)$ component is 
$\alpha_i \beta_j$. This equation can be solved for $\vec{a}$
without determining $\sigma^2$.

\begin{widetext}
The equation for $\sigma^2$ is modified as
\begin{equation}
 \sigma^2
  =
  \frac{1}{M}
  \sum_i
  \Omega_{\text{LOO},i} \nonumber
  \left(
      E[\vec{x}^\text{\,R}_i]
      -
      E'[\vec{x}^\text{\,R}_i]
      -
      \frac{
      \sum_{k \neq i}
      \omega_k
      \left\{
          E[\vec{x}^\text{\,R}_k]
          -
          E'[\vec{x}^\text{\,R}_k]
          +
          \vec{a}
          \cdot
          \left(
              \vec{x}^\text{\,R}_i
	      -
	      \vec{x}^\text{\,R}_k
          \right)
      \right\}
      }
      {\Omega_{\text{LOO},i}}
  \right)^2.
\end{equation}
\end{widetext}

\section{Dimensions for R and Z}
\begin{figure}[htbp]
\centering
 \includegraphics[bb=0 0 504 360, width=7cm]{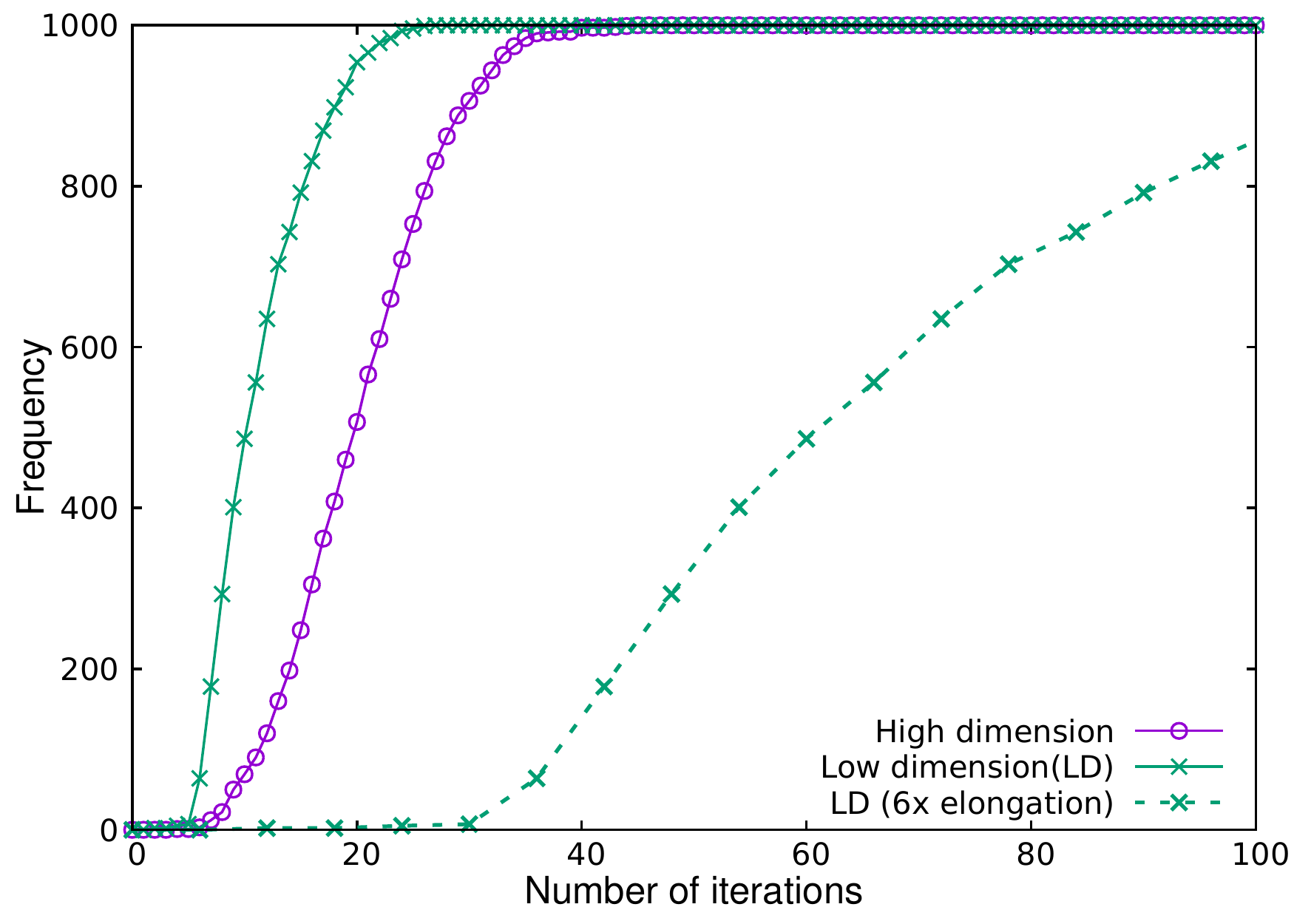}
 \includegraphics[bb=0 0 504 360, width=7cm]{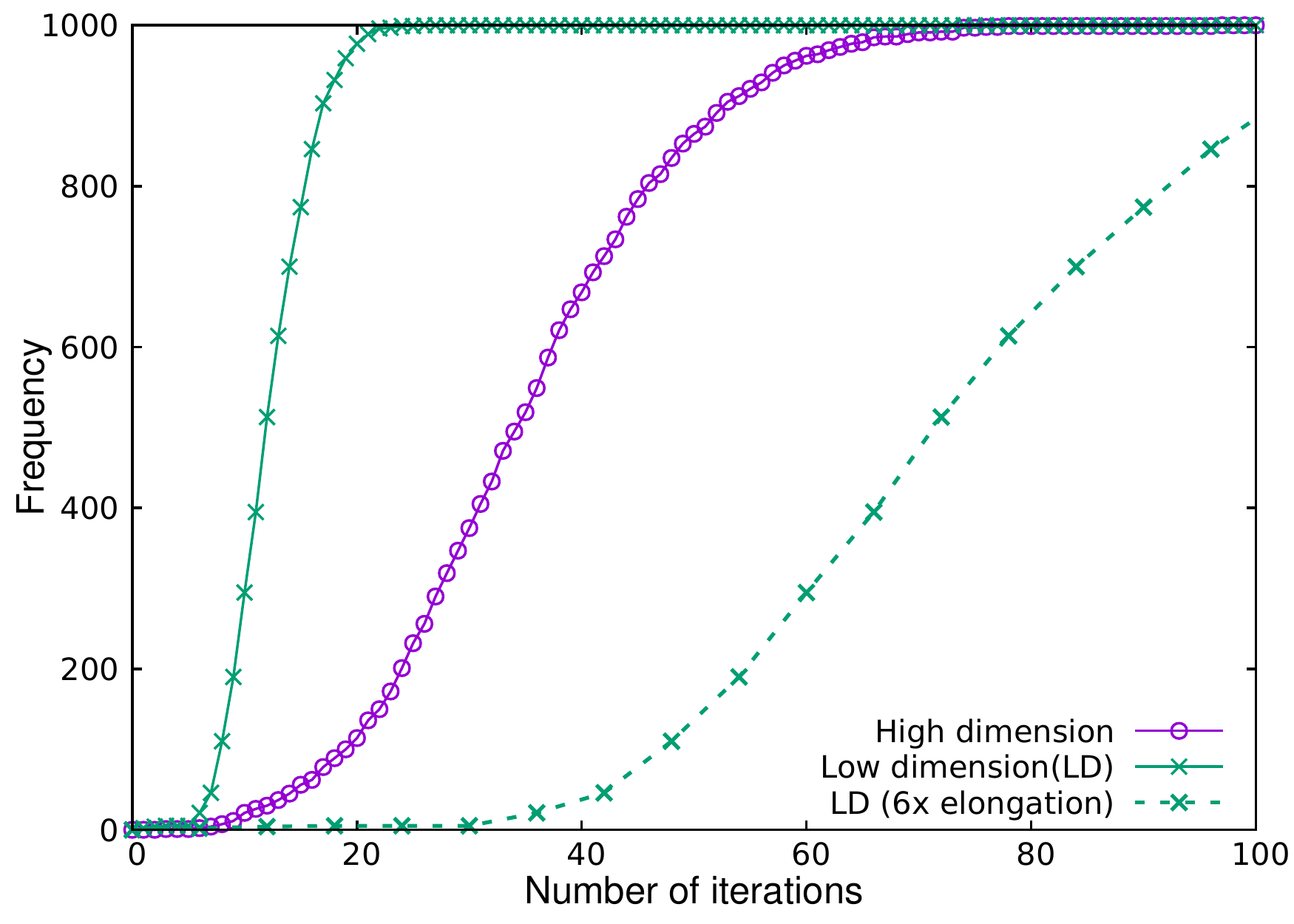}
 \includegraphics[bb=0 0 504 360, width=7cm]{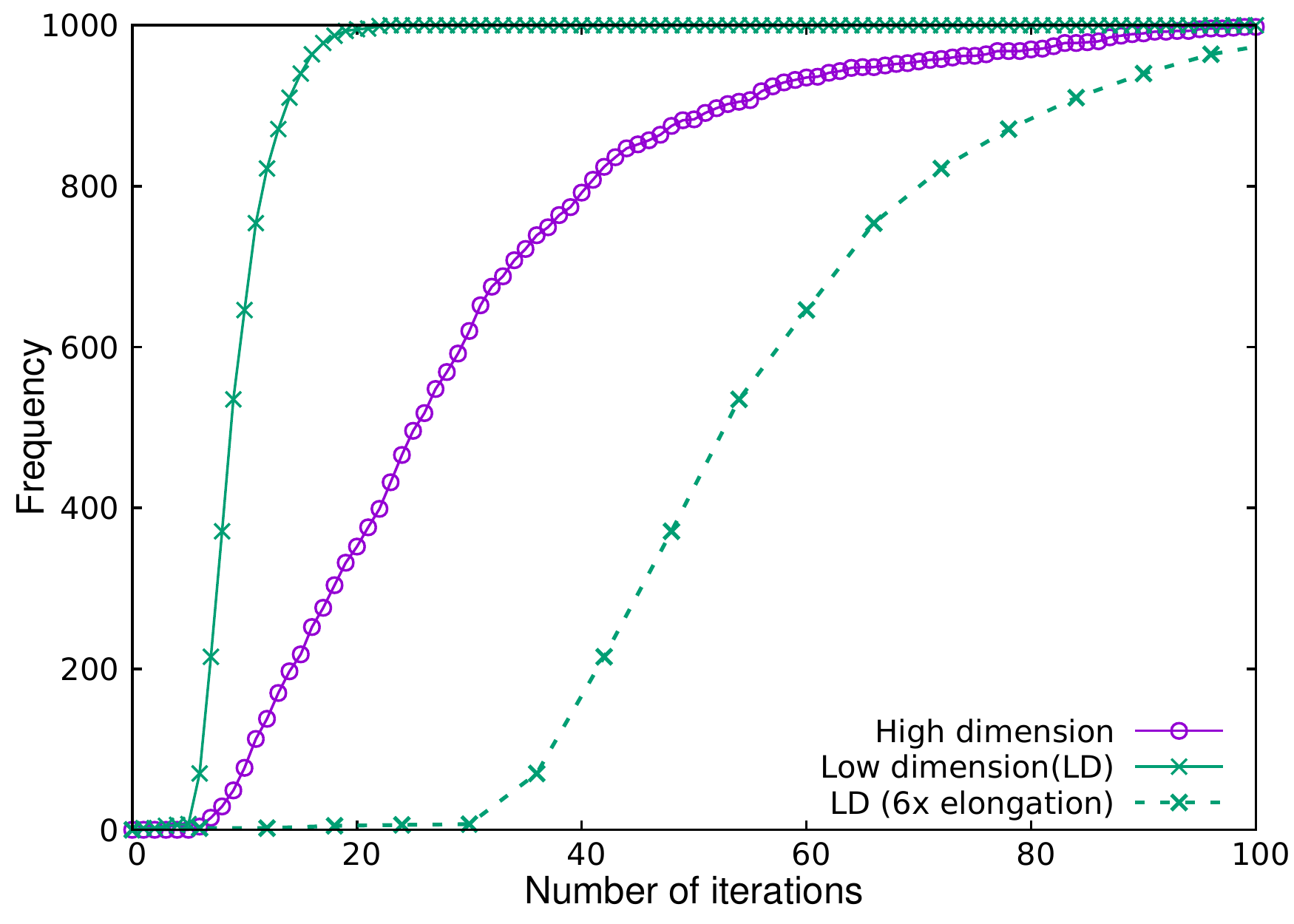}
 \caption{
 Frequency of finding the system
 with the highest magnetization (top),
 Curie temperature (middle), and
 the lowest formation energy (bottom) among 1000 sessions.
 Those 
 in the search for the small space
 where R and Z are fixed to R = Y and Z = Zr
 are denoted by the solid line along cross symbols.
 The dotted line along cross symbols show the curve
 elongated 6 times along $x$-axis.
 The line along the circle symbols show the result
 of the search in the full space using the descriptor \#9
 in Table \ref{tab_descriptors}.
 \label{3vs5}
}
\end{figure}

We prepared dimensions to include information of
the R and Z elements in our design of the descriptors
in Table \ref{tab_descriptors}.
Because the corresponding choice
is only 6 in combination (R = Y, Nd, Sm; Z = Zr, Du)
while it is known that high-dimensionality often causes
problems in modeling,
readers may doubt its necessity.
We compare the efficiency of search for the smaller
space in which 
those elements are fixed to R = Y and Z = Zr
in Fig. \ref{3vs5}.
The solid line along the cross symbols in the figures denote the frequency
of finding the system 
with the best score
out of 1000 sessions
in the restricted space
when we use the set of $\alpha, \beta$ and $\gamma$ as a descriptor.
We also show the curve elongated 6 times along $x$-axis
(the dotted line) because
one has to optimize also for the other combinations of R and Z
to obtain the optimal system for the full space,
which approximately takes 6 times larger.
For comparison, we show the result of optimization
for the full space using the descriptor \#9
by the line along the circle points.
We set the number of iterations before
Bayesian optimization to 5, which is smaller
than the number we use above (=10),
because we know that 
the full-space search
is so fast that the search
in the smaller space cannot catch up
if it starts 
60 (=$10 \times 6$)
iterations behind (Fig. \ref{fig_top10_50} and \ref{Fig_score_vs_steps}).
We see from Fig.\ref{3vs5} that
the search with the full space is more efficient,
even with this setup,
than repeating the search for the small space 6 times.
Therefore, the dimensions for R and Z actually contribute
to the search efficiency.

\section{Data from the first-principles calculations}
\label{rel2exp}
In this section, we list systems with
the predicted highest values of
magnetization and Curie temperature
in order to make a comparison with available
experimental data and to serve a guiding information
for future experimental synthesis.

In Table \ref{magnetization_list}, we show
the top 10 systems of the highest magnetization.
Here we add the contribution of the f-electrons
of Nd, Sm and Dy with the assumption that those have 
$g_JJ$ of 3.27, 0.71 and $-10$ $\mu_B$.
Due to this additional magnetic moment, 
the systems with Nd has advantage over the other 
systems, and all the top 10 systems have 
Nd in them.

It should also be noted that the values for 
the formation energy are positive for all the top 10 systems,
and violate a necessary condition for the thermodynamic stability.
The highest magnetization among the systems
with a negative value of the formation energy is 
obtained by Nd(Fe$_{0.7}$Co$_{0.3}$)$_{12}$.
Note also that this does not ensure the stability
against the other phases.
We also show the values for
(Sm$_{0.7}$Zr$_{0.3}$)(Fe$_{0.9}$Co$_{0.1}$)$_{12}$
because the magnetic anisotropy of Nd tend not to be uniaxial
in ThMn$_{12}$-type systems in the absence of 
a third element.

The best system, NdFe$_{12}$, has already been synthesized by 
Hirayama et al.\cite{Hirayama15}
From the results of our calculations, this system
seems to be near the upper limit of the magnetization
at absolute zero. 
\begin{table}[htbp]
  \caption{
  The top 10 systems with the highest values of the magnetization. 
  The magnetic moment from the f-electrons is assumed to 
  be 3.27 $\mu_{\mathrm B}$ for Nd,
  0.71 $\mu_{\mathrm B}$ for Sm,
  $-10$ $\mu_{\mathrm B}$ for Dy.
  The variable $\mu_0 M$ denotes the magnetization,
  $T_{\mathrm C}$ the Curie temperature, and 
  $\Delta E$ the formation energy from the unary systems.
  The values for Nd(Fe$_{0.7}$Co$_{0.3}$)$_{12}$, 
 which has the highest magnetization among the systems 
 with a negative formation energy, and 
 the values for 
 (Sm$_{0.7}$Zr$_{0.3}$)(Fe$_{0.9}$Co$_{0.1}$)$_{12}$,
 which has the highest magnetization among the systems
 having Sm in them
 with a negative formation energy,
 are also shown.
  \label{magnetization_list}}
 \begin{tabular}{cccc}
  \hline
  \hline
   Formula & $\mu_0 M$ [T] & $T_{\mathrm C}$ [K] & $\Delta E$ [eV] \\
  \hline
  NdFe$_{12}$                     &  1.95 &  844 & 0.405 \\
  Nd(Fe$_{0.9}$Co$_{0.1}$)$_{12}$ &  1.94 & 1012 & 0.230 \\
  Nd(Fe$_{0.8}$Co$_{0.2}$)$_{12}$ &  1.93 & 1111 & 0.095\\
  (Nd$_{0.9}$Zr$_{0.1}$)Fe$_{12}$ &  1.93 &  835 & 0.443\\
  (Nd$_{0.9}$Zr$_{0.1}$)(Fe$_{0.9}$Co$_{0.1}$)$_{12}$ &  1.92 &
	  1011& 0.272 \\
  (Nd$_{0.9}$Zr$_{0.1}$)(Fe$_{0.8}$Co$_{0.2}$)$_{12}$ &  1.91 &
	  1098& 0.140 \\
  (Nd$_{0.8}$Zr$_{0.2}$)Fe$_{12}$ &  1.91 &  841 & 0.446 \\
  (Nd$_{0.8}$Zr$_{0.2}$)(Fe$_{0.9}$Co$_{0.1}$)$_{12}$ &  1.91 &
	  1009& 0.272 \\
  (Nd$_{0.7}$Zr$_{0.3}$)Fe$_{12}$ &  1.89 &  845 & 0.385 \\
  (Nd$_{0.8}$Zr$_{0.2}$)(Fe$_{0.8}$Co$_{0.2}$)$_{12}$ &  1.89 &
	  1086& 0.143 \\
  \hline
  Nd(Fe$_{0.7}$Co$_{0.3}$)$_{12}$ &  1.89 & 1143 & -0.142 \\
  (Sm$_{0.7}$Zr$_{0.3}$)(Fe$_{0.9}$Co$_{0.1}$)$_{12}$ & 1.77 & 1002 &
	      -0.008 \\
  \hline
  \hline
 \end{tabular}
\end{table}

In Table \ref{tc_list}, we show
the top 10 systems with the highest values of 
Curie temperature.
Although the formation energy is negative for all the 
systems in the table, it should be noted again that
those incorporate only the competition with the unary phases
and does not ensure the stability against
the other phases.

The best system in the list is
Sm(Fe$_{0.2}$Co$_{0.8}$)$_{12}$.
Although Hirayama et al. have reported that
they could synthesize Sm(Fe$_{0.8}$Co$_{0.2}$)$_{12}$
as a film,
there is no experimental report for a successful synthesis of
compounds with higher concentration of Co to our knowledge.
\begin{table}[htbp]
  \caption{
  The top 10 systems with the highest values of the Curie temperature. 
  The variable $\mu_0 M$ denotes the magnetization,
  $T_{\mathrm C}$ the Curie temperature, and 
  $\Delta E$ the formation energy from the unary systems.
  \label{tc_list}}
 \begin{tabular}{cccc}
  \hline
  \hline
   Formula & $T_{\mathrm C}$ [K] & $\mu_0 M$ [T] & $\Delta E$ [eV] \\
  \hline
  Sm(Fe$_{0.2}$Co$_{0.8}$)$_{12}$ &  1310 & 1.47 & -0.631 \\
  (Sm$_{0.9}$Dy$_{0.1}$)(Fe$_{0.2}$Co$_{0.8}$)$_{12}$&1309&1.39&-0.654\\
  (Sm$_{0.8}$Dy$_{0.2}$)(Fe$_{0.2}$Co$_{0.8}$)$_{12}$&1307&1.32&-0.676\\
  (Sm$_{0.7}$Dy$_{0.3}$)(Fe$_{0.2}$Co$_{0.8}$)$_{12}$&1305&1.24&-0.696\\
  (Sm$_{0.6}$Dy$_{0.4}$)(Fe$_{0.2}$Co$_{0.8}$)$_{12}$&1304&1.17&-0.715\\
  (Sm$_{0.5}$Dy$_{0.5}$)(Fe$_{0.2}$Co$_{0.8}$)$_{12}$&1302&1.09&-0.733\\
  (Sm$_{0.4}$Dy$_{0.6}$)(Fe$_{0.2}$Co$_{0.8}$)$_{12}$&1300&1.01&-0.752\\
  (Nd$_{0.7}$Dy$_{0.3}$)(Fe$_{0.2}$Co$_{0.8}$)$_{12}$&1300&1.37&-0.553\\
  (Nd$_{0.6}$Dy$_{0.4}$)(Fe$_{0.2}$Co$_{0.8}$)$_{12}$&1299&1.27&-0.594\\
  (Nd$_{0.8}$Dy$_{0.2}$)(Fe$_{0.2}$Co$_{0.8}$)$_{12}$&1299&1.46&-0.510\\
  \hline
  \hline
 \end{tabular}
\end{table}

\bibliography{mito}
\end{document}